\documentstyle[prc,preprint,aps]{revtex}
\begin{document}
\draft
\title{$^{12}$C(p,n)$^{12}$N reaction at 135 MeV} 
\author {B.\ D.\ Anderson,$^1$ L.\ A.\ C.\ Garcia,$^{1,}$\cite{faz} 
D.\ J.\ Millener,$^2$ D.\ M.\ Manley,$^1$ A.\ R.\ Baldwin,$^1$ A.\ 
Fazely,$^{1,}$\cite{faz} R.\ Madey,$^{1,4}$ N.\ 
Tamimi,$^{1,}$\cite{tam} J.\ W.\ Watson,$^1$ and C.\ C.\ Foster$^3$}
\address{$^1$Department of Physics and Center for Nuclear Research,
Kent State University,\\ Kent, Ohio 44242}
\address{$^2$Brookhaven National Laboratory, Upton, New York 11973}
\address{$^3$Indiana University Cyclotron Facility, Bloomington, 
Indiana 47405}
\address{$^4$Hampton University, Hampton, Virginia 23668}
\date{\today}
\maketitle
\begin{abstract}

We report observations from the (p,n) reaction on $^{12}$C 
at 135 MeV. The experiment was performed with the beam-swinger 
neutron time-of-flight system at the Indiana University Cyclotron 
Facility. Neutrons were detected in large-volume plastic 
scintillation detectors located in three detector stations at 
0$^{\circ}$, 24$^{\circ}$, and 45$^{\circ}$ with respect to the 
undeflected beam line; the flight paths were 91 m, 91 m, and 74 m,
respectively.  Overall time resolutions of about 825 ps provided 
energy resolutions of about 350 keV in the first two stations 
and about 425 keV in the third station.  The angular 
distributions for states with excitation energies up to 10
MeV are presented and comparisons are made with DWIA calculations
that use one-body density matrices from $0\hbar\omega$ and 
$1\hbar\omega$ shell-model calculations. New information is 
deduced on the excitation energies, widths and spin-parity 
assignments for several energy levels of $^{12}$N.
\end{abstract}
\pacs{PACS numbers: 25.40.Kv, 21.10.Hw, 21.10.Jx, 21.60.Cs, 27.20.+n}
\narrowtext
\section{Introduction}
\label{sec:sec1}

 Nucleon-induced charge-exchange reactions provide an extremely useful
probe of isovector excitations in nuclei \cite{RA94}. In part, this is
because cross sections and spin observables for strong transitions at
medium energies are described well by distorted-wave impulse 
approximation (DWIA) calculations. More precisely, single-step charge
exchange appears to be the dominant reaction mechanism above 100 MeV
incident energy, with both $t$-matrix and $G$-matrix interactions 
successfully describing experimental cross sections when the  one-body
transition densities are known \cite{LO87}. Experimentally, the energy
resolution is best at low energies and we have performed a number of
studies using the $(p,n)$ reaction at 135 MeV incident energy
\cite{AN85,AN90,FA82,CH86,AN91,TA92}. These includes studies of
Gamow-Teller strength \cite{AN90,AN91}, stretched states 
\cite{TA92,GA94}, and simple particle-hole excitations in closed-shell
nuclei \cite{AN85,FA82,CH86}.

 In closed-shell nuclei, the predominantly single-step reaction 
mainly excites one-particle--one-hole ($1p1h$) final states.  
Such excitations are relatively easy to describe theoretically 
\cite{TO77,RI80} and comparisons with experimental results can 
provide important tests of nuclear structure models. For example,
the strongest excitations observed in the (p,n) reaction on
the closed-shell nuclei $^{16}$O and $^{40}$Ca \cite{FA82,CH86}
are consistent with the predictions of simple shell-model 
calculations \cite {DO70} in the Tamm-Dancoff approximation 
(TDA), which assumes that the target is a closed core and that
the final states are made up of only $1p1h$ configurations. 
Although these shell-model calculations are able to reproduce 
the relative strengths and the excitation energies fairly well,
the absolute strengths calculated in the DWIA are generally too
large by a factor of two or more. To obtain an understanding of
the absolute strengths, the inclusion of $2p2h$ correlations
in the initial and final states is required. This is done in
the extension of the TDA to the random-phase approximation 
(RPA). The quenching, or enhancement, of strength for 
collective states can be clearly demonstrated in simple but 
realistic schematic models \cite{TO77,RI80}.

 The RPA correlations, and others, can be included in more 
sophisticated shell-model calculations, which avoid the 
violations of the Pauli principle inherent in the RPA. For 
example, in the calculations and analyses of Gareev $et~al.$ 
\cite {GA85} for $^{16}$O, which include selected configurations
up to $3\hbar\omega$, the normalization factors required for the
DWIA calculations are much closer to unity. More recently, 
calculations have been performed for $^{16}$O, which include all
configurations up to $4\hbar\omega$ \cite{HA90,WA92}. In Ref. 
\cite{WA92}, a factor of two quenching with respect to the TDA 
was found for the spin-dipole matrix element between the $^{16}$O
and $^{16}$N ground states. The basis sizes for such shell-model
calculations are typically very large unless a realistic symmetry
scheme can be used to truncate the bases; furthermore, 
consistency problems not present for $0\hbar\omega$ or 
$1\hbar\omega$ calculations should be addressed \cite{WA92,MI92}.

  The situation is very similar for open-shell nuclei, as 
exemplified by our studies of the self-conjugate nuclei 
$^{20}$Ne, $^{24}$Mg, $^{28}$Si, and $^{32}$S \cite{AN91,TA92}. 
Most of the experimental $(p,n)$ spectra and angular distributions
are described reasonably well by large-basis shell-model 
calculations (still at the $0\hbar\omega$ or $1\hbar\omega$ level),
although some specific transitions are described poorly.  As for 
the closed-shell nuclei, the theoretical cross sections typically
need to be renormalized by 10\% to greater than 50\% to agree in
magnitude with experiment. In these cases, the multi-$\hbar\omega$
bases are so large that extended-basis
shell-model calculations have not been performed.

The $^{12}$C(p,n)$^{12}$N reaction that we study in this work 
provides a more realistic example for tests of extended-basis
shell-model calculations. Although such calculations have not 
yet been performed, the basis sizes are comparable to those for
$A=16$ and the calculations should be possible. There have been
many studies of charge-exchange reactions on $^{12}$C and 
references to the older literature may be found in recent papers
devoted to $(n,p)$ \cite{BR91,OL93} and $(p,n)$ \cite{YA93} 
studies. The most prominent peaks in charge-exchange spectra are
due to the $1^+$ ground state, a $2^-$ spin-dipole state at 
$\sim 4$ MeV, and $1^-$ dipole and spin-dipole strength centered
around 7 MeV. The resolution in this work is sufficient to 
exhibit clearly two more peaks and to extract cross sections for 
a number of other states by peak fitting.

 The experimental procedure is given in Sec.\ II. The data 
reduction is described and spectra are presented in Sec.\ III. 
The structure and reaction calculations are described in Sec.\ IV,
where existing information on the positive-parity and 
negative-parity states is summarized and interpreted in terms of 
shell-model calculations with the Cohen-Kurath \cite{CO65} and 
Millener-Kurath \cite{MI75} interactions, respectively. Since 
most of the $^{12}$N states of interest are unbound and possess
substantial proton decay widths, calculated Coulomb energy shifts
and decay widths are used to relate the states of $^{12}$N  to 
those of $^{12}$B, which are better known. A detailed comparison
between the measured angular distributions and theory is
made in Sec.\ V; the results are summarized, and conclusions
drawn, in Sec.\ VI.

\section{Experimental Procedure}

The measurements were performed at the Indiana University 
Cyclotron Facility with the beam-swinger system. The experimental 
arrangement and data reduction procedures were similar to those 
described previously \cite{AN85,AN90}. Neutron kinetic energies 
were measured by the time-of-flight (TOF) technique. A beam of 
135-MeV protons was obtained from the cyclotron in narrow beam 
bursts typically 350 ps long, separated by 133 ns. Neutrons were
detected in three detector stations at 0$^{\circ}$, 24$^{\circ}$,
and 45$^{\circ}$ with respect to the undeflected proton beam. 
The flight paths were 90.9 m, 90.8 m, and 74.4 m ($\pm$0.2 m), 
respectively.  The neutron detectors were rectangular bars of 
fast plastic scintillator 10.2 cm thick. Two separate detectors,
each 1.02 m long by 0.25 m high, were combined for a total 
frontal area of 0.51 m$^{2}$ in the 0$^{\circ}$ station, and 
two detectors, each 1.02 m long by 0.51 m high, were combined
for a total frontal area of 1.04 m$^{2}$ in the 24$^{\circ}$ 
station. The 45$^{\circ}$ station had two detectors, one 1.02 m
long by 0.51 m high and the second 1.02~m long by 1.02~m high, 
for a total frontal area of 1.55~m$^{2}$. Each neutron detector 
had tapered Plexiglass light pipes attached on the two ends of
the scintillator bar, coupled to 12.8-cm diameter phototubes. 
Timing  signals were derived from each end and combined in a 
mean-timer circuit \cite{BA80} to provide the timing  signal 
from each detector.  Overall time resolutions of about 825 ps 
were obtained, including contributions from the beam burst width
(350 ps) and energy spread  (480 ps), energy loss in the target
(300 ps), neutron transit times across the 10.2 cm thickness of
the detectors (550 ps), and the intrinsic time dispersion of 
each detector (300 ps). This overall time resolution provided 
an energy resolution of about 350 keV in the first two detector
stations and about 480 keV in the widest-angle station. The 
large-volume detectors were described in more detail previously 
\cite{MA83}. Protons from the target were rejected by 
anticoincidence detectors in front of each neutron detector 
array. Cosmic rays were vetoed by anticoincidence detectors 
on top as well as the ones at the front of each array.

The $^{12}$C target was a 31.4 mg/cm$^{2}$ natural target 
(98.9\% $^{12}$C). The TOF spectra were obtained at 12 angles 
between 0$^{\circ}$ and 63$^{\circ}$. Spectra from each detector
were recorded at many pulse-height thresholds from 25 to 90 MeV 
equivalent-electron energy (MeVee). Calibration of the pulse-height 
response of each of the detectors was performed with a $^{228}$Th 
source (which emits a 2.61-MeV gamma ray) and a calibrated
fast amplifier. The values of the cross sections extracted for 
different thresholds were found to be the same within statistics.  
The values of the cross sections reported are at a threshold 
setting of 40 MeVee.

\section{Data Reduction}

The experimental procedure and data reduction are similar to 
those described in more detail in Refs. \cite{AN85} and \cite{AN90}.  
Excitation-energy spectra were obtained from the measured 
TOF spectra using the known flight paths and a calibration of the 
time-to-amplitude converter. Known states in the residual nucleus 
$^{12}$N provided absolute reference points. Excitation energies 
are estimated to be accurate to 0.1 MeV or better; for example,
to 50 keV for several of the peaks listed in Table I. The 
excitation-energy spectra for five angles at roughly $12^\circ$ 
intervals are presented in Fig.~\ref{fig:spec}.

Yields for individual transitions were obtained by fitting peaks
in the TOF spectra. The spectra were fitted with an improved 
version of the peak-fitting code of Bevington \cite{BE69}. 
Because the proton threshold in the residual nucleus, $^{12}$N,
is at 0.60 MeV, all the final states are unbound except for the
g.s.; consequently, we fit the (p,n) spectra using Lorentzian 
line shapes folded together with a Gaussian line shape to 
account for the experimental resolution. The Gaussian width was 
determined from the fit to the g.s. peak in each spectrum. We set 
the widths of the Lorentzians to be the widths accepted in the 
compilation of Ajzengerg-Selove \cite{AJ90}, except for the 
strong, broad 2$^{-}$, 4$^{-}$ complex at 4.3 MeV and the $3^+$,
$3^-$ peak at 5.4 MeV, which were fitted here to obtain new 
values, as discussed below. The fits included a cubic polynomial 
background that provided a shape very much like that of calculated 
quasi-free scattering [i.e., (p,pn)] spectra as  presented 
previously \cite{AN81,AN85b}. Examples of the fitting are shown
in Fig.~\ref{fig:spec} . Note that a small ``tail" is observed 
on the large g.s. peak, which is fitted with an additional 
Gaussian (see the two forward-angle spectra in 
Fig.~\ref{fig:spec}). Such tails are commonly observed in neutron 
TOF spectra and arise from time-slewing of lower pulse-height 
events. The area of this tail is $\sim$3\% of the total peak 
area and is included in the area for the peak. Such tails cannot 
be observed on the broader unbound states. 
In general, the fits were 
judged to be good. There always remains the question of the 
background under peaks in the continuum region. This uncertainty
affects primarily the high-lying states above 6 MeV.
The results we present here represent a lower limit for these
levels because we are not considering contributions from the 
underlying continuum, such as one might obtain in a ``multipole
analysis" of the entire spectrum, for example. 

We allowed the excitation energy and the Lorentzian width for 
the 2$^{-}$, 4$^{-}$ complex to vary because we observed that 
both changed in a systematic way from forward angles to backward 
angles as the dominant level changed from the 2$^{-}$ to the 
4$^{-}$ state. These levels are broad ($\sim$800 keV) and excited
strongly so that we could determine, with the experimental 
resolution of 350 keV in this work, the energies and the widths 
for both levels; in addition, we could extract an excitation 
energy and width for the 5.4-MeV peak from the two spectra, at 
$24^\circ$ and $30^\circ$, in which the peak is most prominent.
The excitation energies and the widths of states observed 
in this work are compared with the compilation values in Table I.

Cross sections were obtained by combining the yields with the 
measured geometrical parameters, the beam integration, and the
target thickness. Neutron efficiencies were obtained from a 
Monte Carlo computer code \cite{CE79}, which was tested 
extensively at these energies \cite{WA83,AU84}. The uncertainty
in the cross section is dominated by the uncertainty in the 
detector efficiencies, which is estimated to be $12\%$. 
Uncertainties shown in the angular distributions are the fitting 
and statistical uncertainties only.

 Excitation-energy spectra for the $^{12}$C(p,n)$^{12}$N 
reaction at 135 MeV at 0$^{\circ}$, 12$^{\circ}$, 24$^{\circ}$, 
36$^{\circ}$, and 45$^{\circ}$ are shown in Fig.~\ref{fig:spec}.
The strongest transitions are labeled by the $J^{\pi}$ of the 
residual state in $^{12}$N.  For some of the states, the 
$J^{\pi}$ assignments are known from earlier work; for the 
other cases, the identifications were made here by comparing 
the extracted angular distributions with DWIA calculations and
with known analog states in $^{12}$C and $^{12}$B, as described
below. The $J^{\pi}$ assignments for states up to about 6 MeV in
excitation energy are known quite well in the analog $T=1$ nucleus,
$^{12}$B, and are listed in the compilation \cite{AJ90}. The 
analogs of most of these states in $^{12}$C are known also and 
are listed in the compilation.  

\section{Structure and reaction calculations}

 In $^{12}$N, the proton threshold is at 0.601 MeV and thus all
levels except the ground state are particle unstable. The states
of interest for our experiment, including the broad peak centered
around 7 MeV, lie below the $\alpha$ threshold at 8 MeV. The
proton decay widths of these states are generally quite large and
are, in themselves, a useful test of nuclear structure models. The
tabulated widths \cite{AJ90} come mostly from a high-resolution
$^{12}$C$(^3$He$,t)^{12}$N experiment \cite{ST83}, which should
populate the same states as the $^{12}$C$(p,n)^{12}$N reaction.
In $^{12}$B, the neutron-decay threshold is at 3.37 MeV and the
spectrum is quite well-known below 6 MeV in excitation energy.
The analog states in $^{12}$C are known also but a detailed 
interpretation of the spectrum and decay widths is made less 
certain on account of the possibility of isospin mixing with $T=0$
states; isospin mixing is known to exist in several instances and 
is generally expected to be present because states of the same 
space-spin structure but different isospin occur in close proximity
due to an underlying supermultiplet and/or weak-coupling structure.
A comparison of the $T=1$ analog states in $^{12}$B, $^{12}$C, and 
$^{12}$N is presented in Fig.~\ref{fig:levels}. As can be seen, the
$J^{\pi}$ assignments for the low-lying states in $^{12}$N are known
for only about half as many states as are known for the analogs in
$^{12}$B and $^{12}$C. Unless otherwise indicated, if we
state that a certain excitation energy and/or $J^{\pi}$
assignment is ``known'', we will mean that it is listed as such
in the compilation of Ref. \cite{AJ90}. 
 
 The essential features of the structure of the states in 
Fig.~\ref{fig:levels} can be understood from $0\hbar\omega$ and 
$1\hbar\omega$ shell-model calculations for the positive-parity and
negative-parity states, respectively. Such calculations have been 
described, and the one-body density-matrix elements (OBDME) required
for reaction calculations tabulated, in connection with previous 
analyses of inelastic scattering and charge-exchange experiments on
$^{12}$C \cite{BR91,HI84}. In the following subsections, we give 
brief discussions of the spin-parity assignments, first for the 
positive-parity states and then for the negative-parity states. 
Next, we make estimates of the Coulomb-energy differences between
negative-parity states of $^{12}$B and $^{12}$N to predict the 
excitation energies of states in $^{12}$N from the known states in
$^{12}$B. The prediction of particle-decay widths for unbound states
is a by-product of the same calculation. Finally, we specify the 
ingredients of distorted-wave calculations to compute cross sections
using the nuclear structure input.

\subsection{Positive-parity states}

 There are  six positive-parity states below 6 MeV in $^{12}$B with
$J_n^\pi = 1^+_1, 2^+_1, 0^+_1, 2^+_2, 1^+_2$ and $3^+_1$. The
largely $p$-shell character of these states is evident in pickup
reactions. The first five states are observed in the 
$^{13}$C$(d,^3$He$)^{12}$B reaction \cite{MA75} and the $3^+$ state
is clearly seen in the $^{14}$C$(p,^3$He$)^{12}$B reaction 
\cite{AS76}. The $^{12}$C analogs of all six states are excited 
strongly in the $^{15}$N$(p,\alpha)^{12}$C reaction  \cite{MA71} 
in accordance with $p$-shell predictions \cite{KUM75,KU75}. In the
Cohen and Kurath models \cite{CO65}, only the two lowest states 
were included in fits to energy-level data, and the other four 
states cluster between 4.2 and 5.2 MeV in excitation energy for 
all three fitted $p$-shell effective interactions. With the larger
data base on $p$-shell levels now available, similar fits reproduce
the energies of all six levels quite well \cite{WB92}; however, 
changes in the wave functions, and hence the OBDME for inelastic 
scattering, are small (e.g., the $0^+$ state has to have essentially
pure [431] spatial symmetry with $L=1$ and $S=1$). This is evident 
from Table II, which gives the $LS$-coupling OBDME for the CKPOT and
MP4 \cite{DJM} interactions. The $LS$-coupling OBDME can be scaled to
obtain a fit to $(e,e')$ form factors, as was done for the lowest
$1^+$ and $2^+$ states by Brady {\it et al.}~\cite{BR91}, and then
converted to $jj$-coupling and relative coordinates \cite{BR91} for
use with the distorted-wave code code DW81 \cite{DW81}.

 Higher $p$-shell states are predicted to be excited weakly and to 
fall in a region where dipole and  spin-dipole excitations are 
dominant. The lowest $2\hbar\omega$ states, which should be excited
weakly, are also  expected in this region. A rough estimate for the
energy of the lowest $p^6(sd)^2$ $1^+$, $T=1$ state is obtained by 
subtracting the 6.5-MeV energy difference between the lowest $0^+$,
$T=$2 and $1^+$, $T=1$ states in $^{16}$O from the energy of the 
lowest $T=2$ state for $A=12$ (12.75 MeV for $^{12}$B),
which is thought to have a large $2\hbar\omega$ component.
 
\subsection{Negative-parity states}

 For the negative-parity states, we use wave functions computed in 
the full $1\hbar\omega$ space with the Millener-Kurath (MK) 
interaction \cite{MI75}. The general features of this calculation 
have been discussed by Hicks {\it et al.}  \cite{HI84} in connection
with an analysis of magnetic-multipole excitations in $^{12}$C seen
by inelastic electron scattering. In particular, the supermultiplet
symmetry and weak-coupling structure, especially for the $T=1$ states
of interest here, was investigated. An extensive discussion of the
distribution of dipole and spin-dipole strength for these wave 
functions has been given by Brady {\it et al.} \cite{BR91} in 
connection with a study of the $^{12}$C$(n,p)^{12}$B reaction at 
energies around 60 MeV.
The distribution of dipole and spin-dipole strength for the fitted
$p$-$sd$ interactions of Warburton and Brown \cite{WB92} is very
similar \cite{OL93} to that for the MK interaction. The OBDME
necessary for the reaction calculations described in this paper
are listed in Refs.~\cite{BR91,HI84}.

  Because the pickup strength for the removal of $p$-shell nucleons 
from $^{12}$C is exhausted by the lowest two ${3\over 2}^-$ states 
(at 0 and $\sim 5$ MeV) and the lowest ${1\over 2}^-$ state (at 
$\sim 2$ MeV) of the core, a substantial parentage to one or more 
of these states is a prerequisite for the strong inelastic 
excitation of $A=12$ excited states. The $1\hbar\omega$ model 
predicts eight states below 6 MeV in $^{12}$B or $^{12}$N, all of
which have dominant weak-coupling parentages to the ${3\over 2}^-$
ground state or the ${1\over 2}^-$ first-excited state. Experimental
counterparts for seven of these states are known in $^{12}$B. Only a
$0^-$ level, a member of a ${1\over 2}^-_1\otimes 1s_{1/2}$ doublet
with the $1^-_2$ level (at 4.30 MeV in $^{12}$B), has not been 
identified. The MK interaction, which successfully reproduces the 
ordering of known $0^-, 1^-$ doublets in this mass region, puts the
$0^-$ level 0.53 MeV below the $1^-$ level. This assignment would 
put the $0^-$ state close to the 3.76-MeV $2^+$ level in $^{12}$B.
A  $0^-$ state, of unknown (and probably mixed) isospin, has been
found at 18.40 MeV in $^{12}$C \cite{SE65}, 0.8 MeV below the 
analog $1^-$ level.

 A deficiency in the energy predictions from the MK interaction is 
that the separation between states with dominant $1s_{1/2}$ and 
$0d_{5/2}$ parentages is about 1 MeV too small, a feature not much
improved in the fits by Warburton and Brown (see Table IV of 
Ref.~\cite{WB92}). Nevertheless, the admixtures between the 
$s_{1/2}$ and $d_{5/2}$ configurations seem to be correct in the 
sense that the very different shapes of the $(e,e')$  form factors
for the two $2^-$  levels are reproduced \cite{HI84}. The $1s$, 
$0d$  admixtures are tested also by the Coulomb energy and 
decay-width calculations discussed in the following subsection.

 Above 6 MeV, dipole and spin-dipole excitations are expected to
dominate the $^{12}$C$(p,n)^{12}$N cross section. These states also
have dominant parentages, mostly $d$-wave, to the low-lying core 
states, including now the ${5\over 2}^-_1$ and ${3\over 2}^-_2$ 
levels. As can be seen from Fig.~\ref{fig:spec}, the only 
discernable peak occurs at about 7 MeV excitation energy for 
angles near the peak of the dipole angular distribution.

\subsection{Coulomb energy shifts}
\label{sec4c}

   In many cases, such as the present one, the low-lying 
abnormal-parity states in $p$-shell nuclei have a very simple 
structure, expressed in terms of an $sd$-shell nucleon, mainly
$1s_{1/2}$ or $0d_{5/2}$, coupled to a few low-lying parent 
states of the core. The single-particle Coulomb energies for 
these orbits depend on the orbit and its binding energy,
which makes the experimental Coulomb energy shifts a sensitive 
test of the wave function; for example,
it is evident from the comparison of analog state energies in 
Fig.~\ref{fig:levels} that there are substantial shifts in
excitation energy across an isospin multiplet, especially for
states with a large $1s_{1/2}$ parentage. Because the structure of
$^{12}$N is rather poorly known, we try to estimate the binding
energy differences between states in $^{12}$B and $^{12}$N.
To do this, we compute  single-particle Coulomb energies 
$\Delta E_C^{sp}$ for each weak-coupling
component and weight them by the shell-model parentages. 

 To obtain $\Delta E_C^{sp}$, the depth
of a Woods-Saxon well is adjusted to reproduce the neutron
separation energy for a given component in $^{12}$B. The Coulomb
potential of a uniformly charged sphere is then added to the
Woods-Saxon well and the proton separation energy for $^{12}$N
is calculated. In the case of unbound states, the complex energy
$E-i\Gamma /2$ at which the scattering function has a pole defines
the resonance energy and single-particle width, and is found using
the code GAMOW \cite{VE82}; for narrow resonances, this method 
agrees with others. The geometry of the Woods-Saxon well sets the 
overall scale of the direct Coulomb energy. The exchange energy
and other small corrections, including charge symmetry breaking, 
which must be included in a first-principles attempt to calculate
Coulomb energy differences, are ignored and effectively subsumed
into the direct Coulomb energy; nevertheless, the  direct Coulomb
energy exhibits the orbit and binding energy dependence of the 
Coulomb energies. The parameters of the Woods-Saxon well are 
$r_0 = 1.26$ fm, $a = 0.60$ fm, and $V_{so}=6$ MeV (12 MeV for 
the code Gamow \cite{VE82}). The Coulomb radius parameter $r_c$ 
is chosen to give the radius of the potential of a uniformly 
charged sphere, $R = \sqrt{5/3}\ \langle r^2 \rangle^{1/2}_{ch}$.
For $^{12}$C, the rms charge radius is 2.472(15) fm 
\cite{DEV87}, which gives $r_c = 1.394$ fm. The
masses in amu (electron masses subtracted from atomic masses) are
$M(^{11}$B$)=11.0066$, $M(^{11}$C$)=11.0081$, $m_n = 1.0087$ and 
$m_p =1.0073$. The $^{11}$B$+n$ and $^{11}$C$+p$ thresholds are at
3.370 and 0.601 MeV, respectively.

 A breakdown of the calculation to predict the excitation energies
of the eight low-lying negative-parity states (including an 
unknown but expected $0^-$ state) in $^{12}$N from those of 
$^{12}$B is given in Table~\ref{table3}. The dominant parentages
are to the ${3\over 2}^-$ ground state and the 2.125-MeV 
${1\over 2}^-$, 4.445-MeV ${5\over 2}^-$, and 5.020-MeV 
${3\over 2}^-$ excited states of $^{11}$B. The corresponding 
excited states of $^{11}$C are at 2.000, 4.319, and 4.804 MeV 
and the downward shifts of 125, 126, and 216 keV, respectively, 
are taken into account in a correction $\Delta E_x$ to the 
averaged single-particle Coulomb energy shift. The remaining 
parentage, denoted by $E_>\otimes d$ in Table~\ref{table3},
is in part necessary to ensure proper elimination of spurious 
center-of-mass states. Some of this parentage is accounted for by
$0s$-hole strength, particularly for the low-spin states, and some
by parentage to $T={3\over 2}$ states. We include the $E_>$ 
strength along with that for the ${3\over 2}_2^-\otimes d$ strength
(for the more deeply bound states, the single-particle Coulomb 
energies are high and not so orbit dependent, and this increased 
Coulomb energy would be partially compensated for by a decrease 
in Coulomb energy of the $p$-shell core states). The $1^-_2$ and 
$2_2^-$ model states have some $1s$ parentage to the $A=11$ ground
state so that we cannot compute a Coulomb energy by our 
single-particle method, although we can obtain an upper limit by
using the calculated  $1s$ Coulomb energy for the most loosely 
bound $1s$ state (e.g., for the $1^-_1$ state). 

  For the four known negative-parity levels of $^{12}$N, the 
predicted excitation energies are in very good agreement with 
experiment, bearing in mind that the energy of the broad $1^-_1$
level is not very well defined. Agreement of  a similar quality is
obtained using the same procedure for a number of other $p$-shell
nuclei, in particular for the positive-parity states below 10 MeV 
in the  neighboring $^{13}$C, $^{13}$N pair, which increases our 
confidence in the predictive power of such calculations. The 
predicted energy shifts cover a substantial range and exhibit 
clearly both the expected orbit and binding energy dependence, 
the latter being most evident for the lowest $2^-$, $1^-$ doublet,
where the Coulomb energy for the less bound $1^-$ state is 
$\sim 350$ keV lower than that for the $2^-$ level. The energy 
shift associated with different core excitation energies is 
important for three states and is necessary to get agreement with
the experimental energy of the $3^-_2$ level. The predicted 
excitation energies for the $2^-_2$ and $4^-_1$ levels bracket 
the energy of the unresolved peak at 4.14 MeV in the 
$^{12}$C$(^3$He$,t)^{12}$N reaction \cite{ST83}, in which a 
centroid shift with angle was noted. This shift is also  evident 
in Fig.~\ref{fig:spec}, and we have fitted the peak with two 
states. Finally, we note that the small predicted Coulomb energy
shift for the $1^-_2$ level leads to a large shift in excitation
energy from 4.3 MeV in $^{12}$B to $\sim 3.5$ MeV in $^{12}$N. 
This shift puts the level near to degeneracy with a level,
possibly the $2^+_2$ level, seen at 3.53 MeV via the 
$^{10}$B$(^3$He$,n)^{12}$N reaction \cite{FU74}.

\subsection{Nucleon decay widths}

 Nucleon  decay widths for unbound negative-parity states in 
$^{12}$B and $^{12}$N can be estimated by taking the 
single-particle widths of resonances in a potential well 
\cite{VE82}, with the depth adjusted to produce a resonance at
the decay energy for  the neutron or  proton, and multiplying
these by the shell-model spectroscopic factors given in 
Table III. This method will not work for $s$-wave neutron decay
of $^{12}$B or when the decay energy is too high for a 
well-defined single-particle resonance to  exist, as is the case
(noted in Table IV) for the $s$-wave ground-state decays of the 
$1^-_2$ and $2^-_2$ levels of $^{12}$N; for the $\sim 4.1$-MeV 
$2^-$ state, in particular, $s$-wave $p_0$ decay is probably a 
major contributor to the width. Aside from these limitations, 
it can be seen from Table IV that there
is generally good agreement between the calculated widths and the
experimental values. Some small contributions to the widths, such
as $d$-wave competition to dominant $s$-wave decay or small $p_1$
branches, have been omitted from Table IV.

 It is also of interest to look at the structure and widths of the
higher states that give rise to the dipole and spin-dipole strength
centered around 7 MeV in $^{12}$N. Parentage decompositions for
the $1^-_3$, $1^-_4$, $1^-_5$, $2^-_3$ and $2^-_4$ states are given
in Table V. Much of the parentage consists of $d$-wave strength
based on the lowest two states of the core. There is also 
appreciable parentage to the ${5\over 2}^-_1$ and ${3\over 2}^-_2$ 
states at 4.3 and 4.8 MeV, respectively in $^{11}$C. In the case 
of the $1^-_5$ and  $2^-_4$ states, there is  substantial $s$-wave
parentage. For this reason, these states should be very broad.

  Calculated $d$-wave partial widths for $n_0$ and $n_1$ decay in 
$^{12}$B and $p_0$ and $p_1$ decay in $^{12}$N are given in Table
VI. The excitation energies used in $^{12}$B are taken from the 
shell-model calculation (normalized \cite{BR91} to the known energy
of the $4^-_1$ level), while those in $^{12}$N are obtained from 
a rough estimate using a constant single-particle Coulomb energy 
for unbound $d$ orbits of 2.4 MeV.  While there will also be some
$s$-wave width, the $d$-wave widths of $1-2$ MeV are of the right
magnitude to explain the distribution of dipole strength seen 
in this, and other, charge-exchange reactions.

\subsection{Distorted-wave calculations}

  Angular distributions were calculated in the distorted-wave
impulse-approximation (DWIA) using the code DW81 \cite{DW81}. 
These calculations use the  140-MeV $t$-matrix ${NN}$
interaction as parametrized by Franey and Love \cite{FR85}. The
density-dependent $G$-matrix interaction of Nakayama and Love 
\cite{NAK88}, at the same energy, has also been used.
The optical-model parameters are interpolated from the work of 
Comfort and Karp \cite{CO80A}. 

  The nuclear  structure input is taken from the $0\hbar\omega$
and $1\hbar\omega$ shell-model calculations described in the 
previous subsections. Core-polarization corrections, which
take into account the effect of configurations not included in
the model space, are expected to be substantial and to lead to a
multipole-dependent quenching of cross sections for the isovector
transitions of interest. Such effects follow from general 
properties of the effective $NN$ interaction, as demonstrated
in schematic models, perturbative mixing calculations and 
large-basis shell-model calculations; for example, the inclusion
of $p^2\to (sd)^2$ excitations leads to substantial quenching in
dipole and spin-dipole transitions. Often such effects are 
included empirically by scaling selected OBDME to fit electron 
scattering form factors (for analog states). This scaling is best
done in an $LS$ representation; longitudinal form factors for 
normal-parity excitations are related to $\Delta S =0$ OBDME, 
while transverse form factors are usually controlled by 
$\Delta S =1$ OBDME. Also, core-polarization corrections can 
change the shapes of form factors (transition densities), 
particularly at high momentum transfers, and this effect
is sometimes mocked-up by changing the radial scale of the
single-particle wave functions.  Details are discussed on a
state-by-state basis in the next section.  Remaining discrepancies
in the resultant $(p,n)$ cross sections are exhibited by 
normalizing the angular distribution obtained from the 
DWIA calculation to the experimental angular distribution in the
region of momentum transfer corresponding to the $\Delta L$ transfer
where the cross section is maximum (see Figs. 6 through 13 in Ref.
\cite{BR91} for the cross sections corresponding to pure $\Delta L$,
$\Delta S$ excitations).

 The conventional OBDME that result from model calculations,
plus scaling if necessary, are transformed
(essentially  a Talmi-Moshinsky transformation  for  unequal
masses) so that the single-particle  wave functions are expressed
in terms of the relative coordinate between the nucleon and the
$A=11$ core \cite{BR91}.  When harmonic oscillator single-particle
wave functions are used, the appropriate oscillator parameter 
is a factor of $\sqrt{A/(A-1)}$ larger than the conventional
oscillator parameter, which takes a value of 1.64 fm to fit
the rms charge radius of $^{12}$C in a $p$-shell model. The more
realistic Woods-Saxon wave functions are explicitly a function of
the relative coordinate. Cross sections calculated with Woods-Saxon
wave functions are generally smaller on account of the lack of
overlap between the deeply bound $p$-shell neutron in the initial
state and the loosely bound, or unbound, proton in the final state
\cite{BR91,OH87}. 

\section{Comparison between experiment and theory}
 
\subsubsection{The $1^{+}$ ground state.}

The angular distribution for the transition to the strongly 
excited 1$^{+}$ ground state of $^{12}$N is shown in 
Fig.~\ref{fig:gs}. This transition is a good example of a 
so-called Gamow-Teller (GT) excitation ($\Delta L=0$, 
$\Delta S=1$) with the (p,n) reaction. Shown also are 120-MeV
$^{12}$C(p,p$^{\prime}$) cross sections to the analog state at
15.11 MeV in $^{12}$C \cite{CO81}, multiplied by a factor of 
two to account for the different isospin couplings in the
projectile subspace. The agreement between the (p,n) and 
(p,p$^{\prime}$) measurements is quite good, especially at forward 
angles, confirming the absolute normalization of these data.

The solid curve in Fig.~\ref{fig:gs} represents a DWIA calculation
with the 140-MeV $t$-matrix, a set of OBDME adjusted to fit
the $(e,e')$ form factor of the 15.11-MeV level of $^{12}$C (third
line  of Table IV in Ref.~\cite{BR91}) and an oscillator parameter 
$b_{rel}=1.9$ fm from the same fit. At small angles, the DWIA 
calculation agrees quite well (to within 10\%) with the 
experimental angular distribution. In the region of the shoulder 
around $q = 1.3$ fm$^{-1}$, the DWIA calculation substantially 
overpredicts the cross section. The density-dependent $G$-matrix
interaction of Nakayama and Love \cite{NAK88} gives a somewhat 
lower cross section in this region but the agreement with the 
data is still not good, in analogy to the findings of Bauhoff 
{\it et al.} \cite{BA83} in an analysis of $^{12}$C$(p,p')$ data
at 135 MeV. The problems are similar in the $(p,n)$ reaction at 
160 MeV \cite{RA87} and in the $(p,p^{\prime}$) reaction at 
200 MeV \cite{CO82}.

 The $\Delta J =1$ cross sections involve a delicate interplay of
$L=0$ and 2, or equivalently longitudinal and transverse, 
spin-dipole transition densities and interaction components 
\cite{RA87}. To some extent, the $L=0$ and $L=2$ densities can 
be adjusted to fit the corresponding $(e,e')$ form factors 
\cite{BR91} up to $\sim 1.8$ fm$^{-1}$ . The $L=0$ density 
controls the low-$q$ behavior or the BGT value (remembering 
that meson-exchange-current corrections are different for the 
two processes) and the $L=2$ density can be adjusted to 
reproduce the minimum of the  form factor. These effects have been
obtained in core-polarization calculations \cite{SUZ81}.
At larger momentum transfers, no $p$-shell model can reproduce the 
$(e,e')$ form factor (transition density), and core-polarization
calculations do little better, so that DWIA calculations using 
these transition densities cannot be expected
to reproduce the $(p,n)$ or $(p,p')$ cross sections.

\subsubsection{The 2$^{+}$, 2$^{-}$ complex at 1.0 MeV.}

The first excited state of $^{12}$N is known to be a 2$^{+}$
level at 0.96 MeV. Its analogs in $^{12}$B and $^{12}$C are   
at 0.95 MeV and 16.11 MeV, respectively. The 2$^{+}$ state 
in $^{12}$N in this experiment is unresolved from a 2$^{-}$ 
state at 1.19 MeV. The analogs of the 2$^{-}$ state are  
at 1.67 MeV and 16.58 MeV in $^{12}$B and $^{12}$C, respectively.
Figure~\ref{fig:ex10} compares the $^{12}$C(p,n) 
angular distribution for this doublet 
with the 120-MeV $^{12}$C(p,p$^{\prime}$) angular distributions 
to the analog states \cite{CO81}, which could be resolved in that
experiment. As for the ground state, the (p,p$^{\prime}$) cross 
sections were multiplied by a factor of two for comparisons here.

 It is clear from Fig.~\ref{fig:ex10} that the $2^+$ state 
dominates the cross section for the 1 MeV peak. In a $p$-shell 
model, the two (of three) important OBDME for the $2^+$ state, 
those with $\Delta L=2$ $\Delta S=0$ and $\Delta L=2$ $\Delta S=1$,
can be scaled to give a good fit to the longitudinal and transverse
$(e,e')$ form factors, respectively, up to $q\sim 1.5$ fm$^{-1}$.
For harmonic oscillator wave functions, the scaling factors for the
CKPOT interaction are 0.50 and 0.84 for the $\Delta S=0$ and 
$\Delta S=1$ OBDME \cite{CO82}. The corresponding factors for 
Woods-Saxon wave functions are 0.577 and 0.915 (see Fig.\ 15 of 
Ref.~\cite{BR91}). Core-polarization calculations do reduce the 
transverse form factor near the peak and give a strong enhancement
at large $q$ \cite{SA85}, as required by the data. The curves in 
Fig.~\ref{fig:ex10} from the DWIA calculations, which 
use the scaled OBDME and an oscillator parameter $b_{rel}=1.71$ fm, 
determined by the rms charge radius, have been scaled down by
a further factor of 0.7. This additional factor is typical of what
has been found in analyses of $(p,p')$ data \cite{NAK88,CO82}. Near
the peak of the cross section, the central and tensor amplitudes are
comparable and the strong constuctive interference between these
amplitudes leads to a slight overshoot of the data. For momentum
transfers beyond the peak, the spin-orbit interaction also plays an 
important role (see Fig.\ 16 of Ref.~\cite{NAK88}).

 Clearly, little can be said from this experiment concerning the 
role of the $2^-$ state; nevertheless, this is a very interesting
transition, for which the dominance of the $(\lambda\ \mu) = 
(2\ 1)\ \Delta L=1\ \Delta S=1$ OBDME \cite{BR91} gives rise to an
$(e,e')$ form factor peaked at high momentum transfer. The $(e,e')$
form factor is reproduced well with a normalization of 0.65 for 
harmonic oscillator wave functions \cite{HI84} (0.71 for the data
of Deutschmann {\it et al.} \cite{DE83}), while very little 
renormalization is required for Woods-Saxon wave functions; 
therefore, it is surprising that the DWIA calculations overestimate
the measured cross section for this state by a factor of more than
five \cite{CO82}. Near the peak of the cross section, the tensor
interaction dominates with some destructive interference from the
central interaction. The magnitude of the peak cross  section is 
quite insensitive to the choice of radial wave functions, although
the position of the peak shifts with changes in radial scale. 
Because the structure of the state gives rise to a dominant 
$\Delta L=1$ amplitude, both longitudinal and transverse components
of the effective interaction contribute. At higher energies 
(800 MeV), the $(e,e')$ and $(p,p')$ normalization factors are more
nearly commensurate \cite{JO94}. At low incident energies (35 and 
40 MeV), the $(p,n)$ cross section is much larger and peaks at low
$q$ ($\sim 0.7$ fm$^{-1}$), where the cross section is very 
sensitive to the choice of radial wave function \cite{OH87}; 
the cross section is reproduced well when the M3Y interaction is 
used with Woods-Saxon wave functions for the loosely-bound 
$\pi 1s_{1/2}$ and $\pi 0d_{5/2}$ orbits. It would be interesting
to have low-$q$ data at the higher bombarding energies. Further 
study of this and related transitions, such as the excitation of 
the ${5\over 2}^+_2$ state of $^{13}$C or $^{13}$N, would be of 
considerable interest. 

\subsubsection{The 1$^{-}$ state at 1.8 MeV.}

The angular distribution for the broad 1$^{-}$ state at 1.8 MeV is
shown in Fig.~\ref{fig:ex18}. This transition is excited weakly 
and was observed only at three forward angles. Its analogs are at
2.62 MeV in $^{12}$B and at 17.23 MeV in $^{12}$C.  The substantial
shift in excitation across the multiplet, and the large width of 
the state in $^{12}$C and $^{12}$N, are consistent with the large
$1s_{1/2}$ parentage to the $A=11$ ground state obtained in the 
shell-model calculations. Although the structure of the $1^-$ 
state is very similar to that of the $2^-$ member of the doublet
at 1.19 MeV, there is a large enough $(\lambda\ \mu)=(1\ 0)$ 
amplitude \cite{BR91} for the cross section to peak at low $q$ 
rather than at high $q$. The shape of the calculated cross section
fits the limited data quite well with a normalization factor of 
0.20 if harmonic oscillator wave functions are used. The peak of
the cross section shifts to lower $q$ when the more spatially 
extended Woods-Saxon wave functions are used and the cross section
is reduced on account of the reduced overlap between initial- and
final-state single-particle wave functions, with the normalization
factor rising to 0.35. Previously, this state was observed only in
the $^{12}$C$(^3$He$,t)^{12}$N reaction \cite{ST83}, where
the forward-peaked angular distribution is consistent with the
$J^\pi =1^-$ assignment.  

  The $(p,n)$ cross section at 135 MeV is largely a measure of the
$(\lambda\ \mu)=(1\ 0)$, $\Delta S=1$ strength. The ground state 
radiative width of the analog state at 17.23 MeV in $^{12}$C is a 
measure of the $(1\ 0)$, $\Delta S=0$ strength and is given as 
$\Gamma_{\gamma_0}\ge 38.3$ eV \cite{AJ90}. This corresponds to 
BC1$\uparrow\ge 0.022$ $e^2$fm$^2$, which is consistent
with the shell-model prediction of 0.038 $e^2$fm$^2$.

\subsubsection{Remaining states below 4.3 MeV.}

   In a high-resolution study with the $(^3$He$,t)$ reaction 
\cite{ST83}, three relatively narrow peaks were observed at 2.45,
3.14, and 3.57 MeV. We do not see the 2.45-MeV $0^+$ state, which
has analogs in $^{12}$B at 2.72 MeV and in $^{12}$C at 17.76 MeV.
There may be a small amount of strength near 2.4 MeV (see 
Fig.~\ref{fig:spec}), but it is too small for us to extract a 
cross section. The predicted peak cross section for this state, 
without any renormalization, is less than 0.03 mb/sr at 
$q\sim 0.75$ fm$^{-1}$; the $p$-shell OBDME is necessarily 
pure $\Delta L=1$, $\Delta S=1$ and the cross section is due 
mainly to the tensor interaction. The other states appear as a
complex seen as a shoulder on the larger complex of states 
centered near 4.3 MeV (see Fig.~\ref{fig:spec}); for this reason,
the extraction of cross sections for these states, which are 
excited weakly, is difficult and sensitive to the choice of 
lineshapes and backgrounds. At forward angles, we find evidence 
for cross section only at 3.5 MeV, and at wider angles only at 
3.2 MeV. 

 The states expected in this region are the analogs of the 
3.39-MeV $3^-$, 3.76-MeV $2^+$, and 4.30-MeV $1^-$ states of 
$^{12}$B (18.35, 18.80, and 19.2 MeV in $^{12}$C), along with
the $0^-$ partner of the $1^-$ state (possibly at 18.40 MeV in 
$^{12}$C). The $3^-$ and $1^-$ states in $^{12}$N are expected,
on the basis of our Coulomb energy calculations, to be
near 3.1 and 3.5 MeV, respectively (see Table III). 

 The cross section that we extract for a state at $\sim 3.2$ MeV
is shown in Fig.~\ref{fig:ex32} and is very small, reaching only
$\sim 0.036$ mb/sr at $q\sim 1$ fm$^{-1}$. The three points do not
seem to be consistent with any reasonable angular distribution. 
The DWIA calculation gives a cross section for the first $3^-$ 
state that is a factor of eight larger than what we extract, even
after taking into account a quenching factor of two for spin 
excitations in the $0/1\hbar\omega$ model spaces. The calculated
cross section at $q\sim 1.2$ fm$^{-1}$ receives comparable
contributions from the central and tensor interactions, with 
constructive interference. The cross section for the predicted 
$0^-$ state is dominated by the tensor interaction, peaking at 
$q\sim 1.5$ fm$^{-1}$, and is also larger than the extracted 
cross section (see Fig.~\ref{fig:ex32}). The $(^3$He$,t)$ 
angular distribution is consistent with the excitation of a 
$3^-$ state, and the fairly large peak cross section of 
$\sim 0.2$ mb/sr is probably due to the substantial non-spin-flip
amplitude for the $3^-_1$ model state, which is favored at the 
low incident energy per nucleon.

 As can be seen from Fig.~\ref{fig:ex35}, the four low-$q$ data 
points would be fitted well by the calculated cross section for the 
$1^-$ state without renormalization. The DWIA cross section for
the second $p$-shell $2^+$ state, which is expected also at about
3.5 MeV, is shown in Fig.~\ref{fig:ex35} with a normalization 
factor of 0.4. This normalization, which takes into account the 
typical factor of two quenching for isovector spin excitations, 
gives a cross section comparable to that derived from Templon's
analysis \cite{TE93} of the region between strong peaks observed
at 18.3 (mainly $2^-$, $T=0$) and 19.4 MeV (mainly $2^-$, $T=1$)
in $^{12}$C$(p,p')$ at 156 MeV. From our data, it is hard to say
anything  definitive about the excitation of the $2^+$ state. 
The peaking of the $(^3$He$,t)$ cross section at small angles 
\cite{ST83} is consistent also with the excitation of a $1^-$ 
state and the cross section at larger angles suggests a weak 
population of the $2^+$ state (in Ref.~\cite{ST83}, a tentative
$1^+$  assignment was discussed, but this seems unlikely given 
the lack of an analog in $^{12}$B).

\subsubsection{The 2$^{-}$, 4$^{-}$ complex at 4.3 MeV.}

 The angular distribution for the complex of states at 4.3 MeV
is shown in Fig.~\ref{fig:ex43}. This complex is known to include
a 2$^{-}$ state and a 4$^{-}$ state. Analogs of these states are
observed at 4.46 MeV and 4.52 MeV in $^{12}$B, and at 19.4 MeV 
and 19.65 MeV in $^{12}$C, respectively (see Fig.~\ref{fig:levels}).
Figure~\ref{fig:ex43} also shows DWIA calculations 
for transitions to the 2$^{-}_2$ state and the 4$^{-}_1$ state.

The overall shape of the complex is reproduced well with
normalization factors of 0.4 and 0.5 for the transitions to the 
2$^{-}$ and 4$^{-}$ states, respectively, if harmonic oscillator
wave functions are used (not shown) and 0.53 and 0.63 for 
Woods-Saxon wave functions (shown). Beyond $q\sim 1.5$ fm$^{-1}$,
the angular distribution is dominated clearly by the 4$^{-}$ 
transition; hence, the normalization factor required for this 
state is not affected strongly by the details of the calculations
for the lower-spin state in the complex; similarly, the $2^-$ 
state dominates at low $q$. This means that it is possible to 
obtain estimates of the excitation energies and widths of the 
$2^-$ and $4^-$ states from analyses of the low-$q$ and high-$q$
data, respectively. In fits using Lorentzian line shapes folded
with a Gaussian resolution function, whose width is taken from 
the ground-state fit, the excitation energies and widths for the
two states are $E_x = 4.18(5)$ MeV, $\Gamma = 836(25)$ keV and 
$E_x = 4.41(5)$ MeV, $\Gamma = 744(25)$ keV (see Table 
\ref{table1}). The forward angle results are in generally good 
agreement with the $(^3$He$,t)$ result of 4.14(10) MeV and 
830(20) keV \cite{ST83}. The peak cross section of 2.2 mb/sr is
somewhat lower than $\sim 3$ mb/sr from a $(p,n)$ measurement 
at 160 MeV \cite{GA84}, 2.8 mb/sr from a $(p,n)$ measurement
at 186 MeV \cite{YA93},  and 2.8 mb/sr from an $(n,p)$
measurement  at 98 MeV \cite{OL93}. Fits using Gaussian lineshapes,
which are not as good as those using Lorentzians, give cross 
sections lower by $\sim 30$\%, excitation energies lower by 
$\sim 100$ keV and slightly different widths. 

 There have been few analyses of the analog 19.6-MeV complex in
$^{12}$C from $(p,p')$ reactions at incident energies close to 
those of the present experiment. The results of Templon at 156
MeV \cite{TE93}, in which Lorentzian lineshapes were used, are
in good agreement with the present results under the assumption
of good isospin for the $2^-$ state in $^{12}$C. The cross 
sections of Comfort {\it et al.} at 200 MeV \cite{CO82}, obtained
from an  analysis using Gaussian lineshapes, are somewhat lower.
The comparison of $(p,n)$ and $(p,p')$ cross sections for the 
$4^-$ state is complicated by the fact that a pair of 
isospin-mixed $4^-$ levels exist within the 19.6-MeV complex. 
This is particularly evident from the comparison of 
$(\pi^+,\pi^{+}{'})$ and $(\pi^-,\pi^{-}{'})$ cross sections 
\cite{MO79}. Likewise, two $4^-$ states at 19.29 and 19.65 MeV are
included with a $2^-$ state at 19.4 MeV in analyses of 400-, 600- 
and 700-MeV $(p,p')$ data \cite{JO94}. The $2^-$ state also 
appears to be isospin mixed with a predominantly $T=0$, $2^-$ 
state at 18.3 MeV.

 The $2^-_2$ model state contains a  large fraction of the
shell-model spin-dipole strength. The corresponding physical 
states are strongly excited at low $q$ in $(e,e')$, $(p,p')$ 
and charge-exchange reactions; however, a substantial quenching 
of the $1\hbar\omega$ shell-model transition density is 
required to give agreement with the experimentally measured 
cross sections, especially if harmonic oscillator 
single-particle wave functions are used in constructing the 
radial transition density. As noted by Brady {\it et al.} 
\cite{BR91}, two physical effects lead to substantial quenching.
First, the reduced overlap between the deeply-bound initial-state
wave functions and the loosely-bound, or unbound, final-state 
wave functions reduces the reaction cross sections. We find a 
reduction of $\sim 25$\% when the unbound final-state wave 
functions are approximated by Woods-Saxon wave functions bound at
100 keV. Second, as expected on the basis of the schematic model,
the inclusion of $p^2\to (sd)^2$ excitations in the shell-model 
bases leads to substantial quenching of isovector dipole and 
spin-dipole excitations; for example, the inclusion of all states
up to $4\hbar\omega$ for $^{16}$O leads to a factor of two 
quenching for the spin-dipole matrix element to the lowest $2^-$,
$T=1$ state \cite{WA92}. No such comprehensive shell-model 
calculations have been reported for $^{12}$C.

 The $4^-$ state carries a large fraction ($\sim 94$\%) of the 
shell-model M4 strength. This strength should be quenched for 
the same reasons as given above, but the backwards-going 
amplitudes from $p^2\to (sd)^2$ admixtures in the $^{12}$C ground
state should be less destructive than they are for the dipole 
and spin-dipole excitations.

\subsubsection{The 3$^{+}$ and $3^-$ states at 5.4 MeV.}

The angular distribution for the peak at 5.4 MeV is shown in 
Fig.~\ref{fig:ex53}. This peak should contain the analogs 
(see Fig.~\ref{fig:levels}) of the 5.61-MeV $3^+_1$ and 5.73-MeV
$3^-_2$ states in $^{12}$B. Candidates for the $^{12}$C analogs
exist at about 20.5 and 20.6 MeV, respectively. The 20.6-MeV 
complex is clearly observed in inelastic scattering reactions on
$^{12}$C, but may also contain $\Delta T=0$ excitations, which
obviate a direct comparison between $(p,p')$ and $(p,n)$ cross
sections. In fact, the strong stripping strength observed at 20.6 
MeV in the $^{11}$B$(d,n)$ reaction \cite{NE83} cannot be 
accounted for by either of the $T=1$ states (from Table III, the
$3^-$ state has very little ground-state parentage), but can be
accounted for by the fourth shell-model $3^-$, $T=0$ state 
predicted at about this energy. On the other hand, the 
transverse $(e,e')$ form factors \cite{HI84} should be mainly
due to $\Delta T=1$ excitations.

 As can be seen from Fig.~\ref{fig:ex53}, the predicted DWIA 
cross sections are comparable for the $3^+_1$ and $3^-_2$ model
states. The summed cross sections give a reasonably good 
reproduction of the data after renormalization by a factor of 
0.25 for each state when harmonic oscillator wave functions are
used. (The summed transverse form factors, with the $E3$ form 
factor being about 2.5 times the $M3$ form factor, overestimate
the $(e,e')$ data by a similar factor \cite{HI84}). The use of 
loosely bound (100 keV) Woods-Saxon wave functions results in 
$(p,n)$ cross sections that are reduced by factors of 0.76 and
0.85 for the $3^+$ and $3^-$ states, respectively. For the $3^+$
state, the tensor interaction is dominant, while for the $3^-$ 
state, the central and tensor amplitudes are comparable with 
strong constructive interference.

 It should be noted that the cross section extracted for the 
5.4-MeV peak is sensitive to the assumed width (and background
subtraction). The width of 180 keV adopted by Ajzenberg-Selove 
\cite{AJ90} is based mainly on the $(^3$He$,t)$ work of 
Sterrenberg {\it et al.} \cite{ST83}, who analyzed the peak as
two states at 5.3 and 5.6 MeV with widths of 180(30) and 120(50)
keV, respectively. Earlier $(^3$He$,t)$ work \cite{MA76} gives a
width of 400(80) keV for a single peak. This is consistent with 
the width (Table I) that we extract from the spectra at the two
angles at which the peak is seen most clearly. 

\subsubsection{The dipole resonance region.}

 The main peak of the giant dipole resonance in $^{12}$C is 
centered at about 22.5 MeV \cite{AJ90}. In charge-exchange 
reactions, corresponding peaks, with widths of roughly $2-3$ MeV,
are centered around 7.7 MeV in $^{12}$B and slightly lower in 
$^{12}$N. The $1\hbar\omega$ shell model predicts that this 
strength is due mainly to the excitation of $1^-$ states, with
the $\Delta S=0$ and $\Delta S=1$ strength being nearly 
coincident in energy (e.g., Fig. 4 of Ref. \cite{BR91}). Some 
$2^-$ strength is predicted at the low-energy side of the main
$1^-$ strength, but the bulk of the $2^-$ spin-dipole strength
is predicted in the 4.3-MeV peak. The $\Delta J$ splitting of 
the spin-dipole strength, due to the spin-orbit interaction, 
puts the $0^-$ strength nearer to 10 MeV.

 There is evidence from heavy-ion induced charge-exchange 
reactions, which selectively populate spin-flip or non-spin-flip
modes, that the dipole and spin-dipole strength in the 
$\sim 7$-MeV peak is indeed essentially coincident in energy 
\cite{VO88,NA91,IC95,BE93,OK94}. Also, in the $(p,n)$ reaction, 
the strength of the 7-MeV peak relative to the 4-MeV peak
(essentially pure $\Delta S=1$) gets progessively weaker as the 
incident energy increases \cite{GA84}, consistent with the 
energy dependence of the spin-independent part of the effective
interaction.

 Some structure is evident in the giant resonance region of 
$^{12}$C; for example, the longitudinal ($\Delta S=0$) strength
observed at 22.0, 23.8, and 25.5 MeV via $(e,e')$ \cite{YA71} 
is consistent with the structure seen in photonuclear reactions
\cite{YA71,AJ90}. There is less structure in the transverse 
response, although a peak is observed at 22.7 MeV 
\cite{YA71,HI84}. The same peaks are seen in $(p,p')$ reactions
with $\Delta L=1$ angular distributions, except that the 23.8-MeV
structure is resolved into two relatively narrow components at 
about 23.5 and 23.9 MeV \cite{TE93,BU77}. In addition, the 
$(\vec{p},\vec{p}')$ reaction has been used to separate 
$\Delta S=0$ and $\Delta S=1$ contributions to the response 
\cite{BA93}. We have chosen to analyze the strong dipole peak 
in our spectra in terms of peaks at 6.4 and 7.3 MeV, with 
widths of 1200 keV, on the basis of structure observed in the 
$^{12}$C$(^3$He$,t)^{12}$N reaction \cite{ST83,HA94}. We have
also included peaks, with the same widths, at 8.2, 9.1, and 
10.0 MeV to account for strength that is apparent in Fig.\ 2
above the fitted background on the high-energy side of the 
main dipole peak. Such a tail is observed in other experiments
\cite{OL93,YA93} and is expected on the basis of shell-model and
RPA calculations \cite{OL93,YA93}. The energies of the 8.2- and 
10.0-MeV peaks coincide roughly with those of structure in the
giant dipole resonance, referred to above, but the widths are 
chosen arbitrarily. Also, strength is observed at 9.9 MeV in 
the $^{12}$C$(^3$He$,t)^{12}$N and $^{12}$C$(^3$He$,tp)^{11}$C 
spectra of Ref.~\cite{HA94}.
  
The angular distributions for all five states are shown in 
Fig.~\ref{fig:gdr}, together with the summed strength for the
entire region. The angular distributions of the 6.4- and 7.3-MeV
states, which are quite similar in shape and magnitude, are 
clearly consistent with the calculated dipole angular 
distributions of the third or fourth $1^-$ states, to which they 
are compared. The angular distributions for the 8.2- and 10.0-MeV
states also appear to be dipole in nature, while the strength 
at 9.1 MeV is rather weak. 

 For reference, the predicted cross sections at $q=0.52$ fm$^{-1}$
for pure dipole, spin-dipole $1^-$ and spin-dipole $2^-$ states, 
using harmonic oscillator wave functions, are 2.97, 12.07, and 
13.38 mb/sr, respectively. The cross sections for the central 
interaction alone are 2.97, 6.43, and 11.63 mb/sr. For the 
tensor interaction alone, the cross sections are 0, 1.06, and
3.11 mb/sr. Thus, there is strong constructive interference
between the central and tensor interactions for the $1^-$ 
spin-dipole state; the effect of the tensor force on the 
angular distribution can be seen by comparing the theoretical
curves for the $1^-_3$ and $1^-_4$ model states in 
Fig.~\ref{fig:gdr}. The $1^-_4$ model state contains more than
half of the dipole and spin-dipole strength predicted in this 
region \cite{BR91}. The $1^-_3$ model state carries most of 
the $\Delta S=0$ dipole strength and has a calculated cross 
section which is one third that of the $1^-_4$ state on account
of the relative weakness of the spin-independent interaction
at 135 MeV. The predicted cross section for the $2^-_3$ and 
$2^-_4$ states, mostly due to the $2^-_4$ state, is about half
that of the $1^-_4$ state. The $1^-_5$ state also contains 
considerable spin-dipole strength \cite{BR91}, which should be
rather broadly distributed because of a large $s$-wave proton 
decay width to the 4.8-MeV ${3\over 2}^-$ state of $^{11}$C 
(see Table V). 

 The near equality in cross section for the 6.4- and 7.3-MeV 
peaks, seen also for the $(^3$He$,t)$ cross sections, suggests
that the peaks contain comparable amounts of dipole and 
spin-dipole strength in contradiction to the detailed 
predictions of the $1\hbar\omega$ shell model. Thus, it is 
more appropriate to compare the summed strength (both absolute
and relative to the $2^-$ spin-dipole strength at 4.2 MeV) to 
model predictions and to that observed in other experiments.
The latter comparison is of particular interest because the 
subtraction of background in the giant resonance region is a 
difficult and not clearly defined procedure. The comparison in
Table VII shows that our summed cross section of $\sim 3.3$ 
mb/sr is slightly lower than that extracted in other $(p,n)$ 
and $(n,p)$ experiments at $100-200$ MeV incident energy. The 
theoretical prediction for the summed $1^-_3$, $1^-_4$, $2^-_3$
and $2^-_4$ model states is 7.9 mb/sr if harmonic oscillator
wave functions are used. We expect that this value would be 
reduced by factor approaching two if more realistic radial wave
functions were used and if an extended shell-model calculation 
to take into account ground-state correlations were performed;
more specifically, the RPA calculations reported in 
Ref.~\cite{YA93} give $\sim {2\over 3}$ the $\Delta L=1$ cross
section of the $(0+1)\hbar\omega$ shell-model calculation and 
our estimate for the ratio of the cross sections for the 
$2^-_2$ state with Woods-Saxon and harmonic oscillator wave 
functions is $\sim 0.75$ (see Sec. V.5).

\section{Summary and Conclusions}

  The $^{12}$C$(p,n)^{12}$N reaction was studied at 135 MeV with 
energy resolutions of 350 to 425 keV. Angular distributions were
extracted for all peaks observed up through the region of the
giant dipole resonance at an excitation energy of around 7 MeV.
In most cases, the peaks are thought to contain contributions from
more than one state and, where possible, peak fitting was carried 
out. The $J^\pi$ assignments for some of the states are already 
known; for the other cases, identifications were made by comparing
the angular distributions with DWIA calculations and by comparing
with the known analog states in $^{12}$C and $^{12}$B. In this 
connection, the Coulomb energy shifts and nucleon decay widths 
were estimated theoretically for negative-parity states using the
known spectrum of $^{12}$B as a starting point, along with the 
shell-model structure of the states.

 In the first few MeV, we see the $1^+$ ground state, an unresolved
complex containing the $2^+$ state at 0.96 MeV and a $2^-$ state at
1.19 MeV, and a weakly excited $1^-$ state at 1.8 MeV. The analogs
of these states are all well known. The $(p,n)$ angular 
distributions agree with the analog $(p,p')$ angular distributions
for the $1^+$ ground state and the $2^+$,$2^-$ complex at 1 MeV. 
We do not see the reported $0^+$ state at 2.44 MeV, but this is 
not surprising because the predicted cross section is very weak.

 From peak fitting, we see evidence for weakly excited states at 
3.2 and 3.5 MeV, which form a shoulder to the strongly excited 
4.3-MeV peak. These states should correspond to states seen clearly
at 3.14 and 3.57 MeV via the $^{12}$C$(^3$He$,t)$ reaction. The 
major contributors to the cross sections for these two peaks are 
most likely the $3^-$ and $1^-$ analogs of states in $^{12}$B at 
3.39 and 4.30 MeV. The analog of the 3.76-MeV $2^+$ state of 
$^{12}$B may contribute also to the 3.5-MeV peak (see 
Fig.~\ref{fig:ex35} ). Between 4.1 and 4.3 MeV, we see the 
$2^-$,$4^-$ complex with known analogs in $^{12}$C and $^{12}$B.
At forward angles, the $2^-$ state is strongest, and at backward
angles, the $4^-$ state dominates. This fact allowed us to obtain
estimates of the excitation energies and widths of the $2^-$ and
$4^-$ states from analyses of the low-$q$ and high-$q$ data.
At 5.4 MeV, we see clearly a peak with an angular distribution 
that could be described by the excitation of either the $3^+$ or
the $3^-$ states known in the analog nuclei, or by a combination
of both (the predicted cross sections are comparable). Between 6
and 8 MeV, we see a broad distribution of strength described well
by a $\Delta L=1$ angular distribution. We have analyzed this 
strength in terms of three peaks, with the two strongest ones, 
at 6.4 and 7.3 MeV, based on a previous analysis of $(^3$He$,t)$
data \cite{ST83}. These states carry a significant fraction of 
the $1^-$ dipole and spin-dipole strength, although there could 
be some $2^-$ strength in this region as well. A major difficulty
in the giant resonance region, as in all such studies, is 
an uncertainty in the background contribution.

 All the states below 6-MeV excitation energy in $^{12}$B, and 
thus $^{12}$N, and the essential features of the dipole and 
spin-dipole strength in the giant resonance region can be 
accounted for by $0\hbar\omega$ shell-model calculations for 
the positive-parity states and by $1\hbar\omega$ calculations
for the negative-parity states. For these restricted model spaces,
there are substantial core-polarization corrections to the 
shell-model transition densities to be used in inelastic 
scattering calculations. A major effect for the isovector 
transitions of interest is a substantial quenching of transition
strength at low momentum transfer for most multipoles. The loose
binding of the final-state single-particle wave functions makes 
it important to use realistic single-particle wave functions, 
although it is difficult to do this precisely for unbound final
states in $^{12}$N; the lack of overlap between initial- and 
final-state wave functions generally leads to substantial 
reductions in cross section compared to those calculated with 
harmonic-oscillator wave functions. In the absence of 
satisfactory multi-$\hbar\omega$ shell-model calculations for 
$A=12$, scaling factors for certain $LS$ OBDME are introduced. 
With a few exceptions, this procedure results in a consistent 
description of $(e,e')$ form factors and $(p,n)$ angular 
distributions up to $q\sim 1.5-2$ fm$^{-1}$; the quenching
factors show a systematic behavior for $p$-shell nuclei and are
in qualitative agreement with perturbative estimates. A notable
exception occurs for the $2^-_1$ level, not resolved in our
$(p,n)$ data but observed in $(p,p')$, where the $(p,n)$ or 
$(p,p')$ cross section is driven by the tensor force. At higher
$q$, the $(e,e')$ and $(p,n)$ cross sections often exceed the 
distorted-wave predictions, a phenomenon clearly evident but 
not well understood for the ground-state transition (the 
excitation of particles to higher orbits by the tensor force 
is known to provide a significant contribution). 

 The good energy resolution of the present experiment has 
enabled us to extract cross-section data for more states than
previous $(p,n)$ experiments. The new states include the broad
$1^-$ state at 1.8 MeV, the $3^-$ state at 3.13 MeV, the second
$1^-$ state at 3.5 (this peak should also contain a contribution
from the second $2^+$ state) and a $3^+$, $3^-$ doublet near 
5.4 MeV; in addition, new information on the excitation energies
and widths of the $2^-$ and $4^-$ members of the 4.3-MeV doublet
has been extracted from the data at momentum transfers where one
or another of the states dominates the cross section. The 
calculated shifts in excitation energy from $^{12}$B to $^{12}$N
for negative-parity states are in good agreement with the data 
for known states of $^{12}$N, and lend strong support to the 
assignment of a $1^-$ state in the 3.5-MeV complex. Likewise,
the calculated proton decay widths for the negative-parity states
are in generally good agreement with the widths extracted from 
$(^3$He$,t)$ data and the present $(p,n)$ data. The result is a 
better understanding of the spectrum of $^{12}$N.

\acknowledgements
We wish to thank the staff at the Indiana University Cyclotron 
Facility for help in mounting and running this experiment. 
This work was supported in part by the National Science Foundation
and by the U.S. Department of Energy under contract
number DE-AC02-76CH00016.

\newpage

\begin{figure} 
\protect\caption{Excitation-energy spectra at $0^\circ$, 
$12^\circ$, $24^{\circ}$, $36^{\circ}$ and $45^{\circ}$ for 
the $^{12}$C(p,n)$^{12}$N reaction at 135 MeV. Fits to the 
neutron time-of-flight spectra are shown.
See text for discussion.\label{fig:spec}}
\end{figure}

\begin{figure} 
\protect\caption{Comparison of the $T=1$ energy 
levels in $^{12}$B, $^{12}$C, and 
$^{12}$N (from this work and Ref.~\protect\onlinecite{AJ90}).
Energies in $^{12}$C have been renormalized by 15.11 MeV
(the excitation energy of the analog of the $^{12}$N and the
$^{12}$B ground states).
\label{fig:levels}}
\end{figure}

\begin{figure} 
\protect\caption{Angular distribution (solid circles) for the 
$^{12}$C(p,n)$^{12}$N reaction at 135 MeV to the 1$^{+}$ ground
state. The open circles are 120-MeV $(p,p^{\prime})$ data 
($\times$2) for the transition to the analog state in $^{12}$C 
\protect\onlinecite{CO81}. The solid line represents a DWIA 
calculation using one-body density-matrix elements fitted to 
the $(e,e')$ form factor of the analog level in $^{12}$C
(see text).\label{fig:gs}}
\end{figure}

\begin{figure} 
\protect\caption[a]{Angular distribution (solid circles) for the 
$^{12}$C(p,n)$^{12}$N reaction at 135 MeV to the 2$^{+}$, 
2$^{-}$ complex at 1.0 MeV.  The open symbols are 120-MeV 
$(p,p')$ data ($\times$2) for the transitions to the analog 
states in $^{12}$C, from Ref. \cite{CO81}. The curves represent
DWIA calculations using harmonic oscillator wave functions with
the normalizations shown.  The transition densities, based on 
the $p$-shell calculation of Cohen and Kurath (CK) 
(Ref.~\protect\onlinecite{CO65}) for the $2^+$ state and the 
$1\hbar\omega$ calculation of Millener and Kurath (MK) 
(Ref.~\cite{MI75,BR91}) for the $2^-$ state, have already
been adjusted to fit electron scattering data in the case of 
the $2^+$ state (see text).\label{fig:ex10}}
\end{figure}

\begin{figure} 
\protect\caption{Angular distribution for the 
$^{12}$C(p,n)$^{12}$N reaction at 135 MeV to the 1$^{-}$ state
at 1.8 MeV. The solid and dashed curves represent DWIA 
calculations for the MK wave function using Woods-Saxon and
harmonic oscillator wave functions, respectively, with the 
normalizations shown. \label{fig:ex18}}
\end{figure}

\begin{figure} 
\protect\caption{Angular distribution for the 
$^{12}$C(p,n)$^{12}$N reaction at 135 MeV to the $3^-$ state 
at 3.2 MeV. The solid curve represents a DWIA calculation for
MK wave function using harmonic oscillator wave functions with
the normalization shown. The dashed curve shows a similar 
calculation for a $0^-$ state expected near this
excitation energy.\label{fig:ex32}}
\end{figure}

\begin{figure} 
\protect\caption{Angular distribution for the 
$^{12}$C(p,n)$^{12}$N reaction at 135 MeV to the 1$^{-}$, 
2$^{+}$ complex at 3.5 MeV.  The curves represent DWIA 
calculations for the MK and CK wave functions, respectively,
using harmonic oscillator wave functions with the 
normalization factors shown.\label{fig:ex35}}
\end{figure}

\begin{figure} 
\protect\caption{Angular distribution for the 
$^{12}$C(p,n)$^{12}$N reaction at 135 MeV to the 2$^{-}$, 
4$^{-}$ complex at 4.3 MeV. The curves represent DWIA 
calculations for the MK wave functions using Woods-Saxon 
wave functions with the $sd$ orbits bound at 100 keV and with 
the normalizations shown. \label{fig:ex43}}
\end{figure}

\begin{figure}
\protect\caption{Angular distribution for the 
$^{12}$C(p,n)$^{12}$N reaction at 135 MeV to the $3^{+}$, 
$3^-$ complex at 5.4 MeV. The curves represent DWIA 
calculations for CK and MK wave functions, respectively,
using harmonic oscillator wave functions with the 
normalization factors shown.\label{fig:ex53}}
\end{figure}

\begin{figure} 
\protect\caption{Angular distribution for the 
$^{12}$C(p,n)$^{12}$N reaction at 135 MeV to the assumed 
1$^{-}$ states at 6.4, 7.3, 8.2, 9.1, and 10.0 MeV. 
The solid and chain-dashed curves represent DWIA 
calculations for the $1^-_4$ MK state with normalizations 
of 0.89 and 0.33 respectively, using Woods-Saxon wave 
functions with $sd$ orbit bound at 100 keV. The dotted
curve represents a DWIA calculation for the $1^-_3$ MK state
with a normalization of 1.05.\label{fig:gdr}}
\end{figure}

\newpage
\begin{table}
\caption{Energy levels of $^{12}$N.}
\begin{tabular}{cccccc}
\multicolumn{3}{c}{This~work} & 
\multicolumn{3}{c}{Ref. \cite{AJ90}} \\
\tableline
$E_{x}$(MeV)&$J^{\pi}$&Width(keV)&$E_{x}$\ (MeV)
&$J^{\pi}$&Width(keV)  \\
\tableline
$0.0$ & 1$^{+}$ & 0	      & $0.0$   & 1$^{+}$& 0   \\
$1.0$ & 2$^{+}$,2$^{-}$ &     & $0.960$ & 2$^{+}$ & $\leq 20$  \\
      &       &		      & $1.191$ & 2$^{-}$ & 118  \\
$1.8$ & 1$^{-}$&              & $1.80$  & 1$^{-}$ & 750  \\
      &       &		      & $2.439$ & 0$^{+}$ & 68  \\
$3.2$ & (3$^{-}$) & 	      & $3.132$ & 2$^{+}$, 3$^{-}$ & 220 \\ 
$3.5$ & (1$^{-}$,2$^{+}$) &   & $3.558$ & (1)$^{+}$ & 220  \\
$4.18(5)$ & 2$^{-}$ & 836(25) & $4.140$ & $2^{-}+4^{-}$ & 825  \\
$4.41(5)$ & 4$^{-}$ & 744(25)	& & &  \\
$5.40(5)$ & 3$^{+}$, 3$^{-}$ & 385(55) & $5.348$ & 3$^{-}$ & 180  \\
$6.4$ & 1$^{-}$ &               & $6.40$ & (1$^{-}$) & 1200  \\
$7.3$ & 1$^{-}$ &               & $7.40$ & (1$^{-}$) & 1200  \\
\end{tabular}
\label{table1}
\end{table}

\begin{table}
\caption{One-body density-matrix elements in $LS$ coupling for the
even-parity transitions. For each state, the first line refers to 
the MP4 interaction and the second to the Cohen and Kurath POT
interaction.}
\begin{tabular}{ccccccc}
$J_n^{\pi}$ & $E_{x}$(MeV)\tablenotemark[1] & $L=0\ S=1$ & 
$L=1\ S=0$ & $L=1\ S=1$ & $L=2\ S=0$ & $L=2\ S=1$ \\
\tableline
$1_{1}^{+}$ & $-$0.53 & ~0.2886 & $-$0.0973 & ~0.7659 & & ~0.1681 \\
            & ~0.03 & ~0.2262 & $-$0.0327 & ~0.7280  &  & ~0.1361 \\
$2_{1}^{+}$ & ~0.91 & & & ~0.3734  & $-$0.4229 & ~0.4008 \\
            & ~1.64 & & & ~0.3712  & $-$0.4479 & ~0.5137 \\
$0_1^+$     & ~2.54 & & & ~0.5578  & & \\
            & ~4.54 & & & ~0.6243  & & \\
$2_{2}^{+}$ & ~3.90 & & & ~0.3598  & ~0.1316 & $-$0.0954 \\
            & ~4.99 & & & ~0.3899  & ~0.1763 & $-$0.1422 \\
$1_{2}^{+}$ & ~4.51 & ~0.0138 & $-$0.3117 & ~0.0646 & & $-$0.2390 \\
            & ~4.45 & ~0.0240 & $-$0.4062 & ~0.1135 & & $-$0.2427 \\
$3_1^+$     & ~5.04 & & & & & ~0.2628 \\
            & ~4.56 & & & & & ~0.2978 \\
\tablenotetext[1]{Theoretical energies relative to the ground state.}
\end{tabular}
\label{table2}
\end{table}

\begin{table}
\caption[a]{Coulomb energy differences between $^{12}$B and 
$^{12}$N for negative-parity states. For each weak-coupling 
configuration indicated in the first column, parentages for the
eight states listed in the first row are given. Single-particle
Coulomb energies (in MeV), calculated using the neutron energy 
for the specified component, are given in the next row.
The parentage not accounted for by the lowest four core states 
is denoted by $E_>$ and is included with the $3/2^-_2\otimes d$
strength in the calculation.  $\Delta E^{th}_C$ is the sum of
single-particle Coulomb energies $\Delta E^{sp}_C$ weighted by 
parentages. $\Delta E_x$ is a similarly computed correction to 
account for the fact that the excitation energies of $^{11}$C 
states are lower than in $^{11}$B. $\Delta E^{sp}_C$ and 
$\Delta E_x$ are used to obtain theoretical excitation energies
for states in $^{12}$N, which are then compared with the 
experimental values, $\Delta E$ being the difference in keV.}
\begin{tabular}{lcccccccc}
  & $2_1^-$ & $1_1^-$ & $3_1^-$ & $0_1^-$ & $1_2^-$ & $2_2^-$
 & $4_1^-$ & $3_2^-$ \\
 $E_x^{exp}(^{12}$B) & 1.674 & 2.621 & 3.389 & 
(3.77)\tablenotemark[1] & 4.301 & 4.460 & 4.518 & 5.726 \\
\tableline
${3\over 2}_1^-\otimes s$ & 0.697 & 0.762 & & & 0.024 & 0.163 & & \\ 
 $\Delta E_C^{sp}$ & 2.105 & 1.687 & & & \tablenotemark[2] & 
 \tablenotemark[1]& & \\
 ${3\over 2}_1^-\otimes d$ & 0.195 & 0.074 & 0.879 & 0.002 & 0.002
 & 0.476 & 0.815  & 0.034 \\ 
 $\Delta E_C^{sp}$ & 2.707 & 2.598 & 2.509 & 2.478 & 2.452 & 2.447
 & 2.445 & 2.426 \\
 ${1\over 2}_1^-\otimes s$    & & 0.032 & & 0.879 & 0.772 & & & \\ 
 $\Delta E_C^{sp}$ & & 2.406 & & 2.117 & 1.893 & & \\
 ${1\over 2}_1^-\otimes d$    & & & 0.004 & & & 0.255 & & 0.698 \\ 
 $\Delta E_C^{sp}$ & & & 2.747 & & & 2.633 & & 2.490 \\
 ${5\over 2}_1^-\otimes s$ & & & 0.042 & & & 0.016 & & 0.108 \\ 
 $\Delta E_C^{sp}$ & & & 2.602 & & & 2.479 & & 2.240 \\
 ${5\over 2}_1^-\otimes d$ & 0.062 & 0.032 & 0.022 & 0.068 & 0.047 
 & 0.022 & 0.099 & 0.076 \\ 
 $\Delta E_C^{sp}$ & 2.954 & 2.919 & 2.887 & 2.869 & 2.842 & 2.833
 & 2.830  & 2.745 \\
 ${3\over 2}_2^-\otimes s$ & & 0.002 & & & 0.013 & & & \\ 
 $\Delta E_C^{sp}$ & & 2.713 & & & 2.568 & & & \\
 ${3\over 2}_2^-\otimes d$ & 0.004 & 0.003 & 0.005 & 0.014 & 0.099
 & 0.004 & 0.022 & 0.048 \\ 
 $E_>\otimes d$ & 0.042 & 0.095 & 0.048 & 0.037 & 0.043 & 0.064
 & 0.064 & 0.036 \\ 
 $\Delta E_C^{sp}$ & 2.973 & 2.941 & 2.911 & 2.895 & 2.871 & 2.863
 & 2.860 & 2.790 \\
 & & & & & & & & \\
 $\Delta E_C^{th}$ & 2.309 & 1.942 & 2.543 & 2.209 & 2.046 & 2.408
 & 2.519 & 2.505 \\
 $\Delta E_x$ & $-$0.017 & $-0.030$ & $-$0.020 & $-$0.129 & 
 $-$0.129 &  $-$0.051 & $-$0.031 & $-$0.128 \\
 $E_x^{th}(^{12}$N) & 1.197 & 1.764 & 3.143 & 3.081 & 3.449 & 4.048
 & 4.237 & 5.334 \\
 $E_x^{exp}(^{12}$N) & 1.191 & 1.8(3) & 3.132 & & & (4.14) & (4.14)
 & 5.348 \\
 $\Delta E$ (keV)    & 6 & $-$36 & 11 & & & & & $-$14 \\
\tablenotetext[1]{Theoretical excitation energy.}
\tablenotetext[2]{Not calculated since the $1s_{1/2}$ neutron in 
$^{12}$B is unbound.}
\end{tabular}
\label{table3}
\end{table}

\begin{table}
\caption[b]{Experimental decay widths $\Gamma^{expt}$ compared 
with predicted neutron decay $\Gamma^{th}$ widths for states in
$^{12}$B and proton decay widths for states in $^{12}$N. 
$\Gamma^{sp}$ is the calculated width of a single-particle 
resonance, for the well geometry of Sec.~\ref{sec4c}, at an 
energy $E_N$ above threshold for the specified decay channel.}
\begin{tabular}{cccccccc}
 & $J^\pi$  & $E_x$ & Decay & $E_N$ & $\Gamma^{sp}$ & 
 $\Gamma^{th}$ & $\Gamma^{expt}$ \\
 & & (MeV) & & (MeV) & (keV) & (keV) & (keV) \\
\hline
 $^{12}$B & $3^-$ & 3.389 & $n_0(d)$ & 0.019 & 0.0034 & 0.0030
 & 0.0031(6) \\
 & $4^-$ & 4.518 & $n_0(d)$ & 1.148 & 145 & 118 & 110(20) \\
 & $3^-$ & 5.726 & $n_0(d)$ & 2.356 & 748 & 25.4 & \\
 &       &       & $n_1(d)$ & 0.231 & 1.6 & 1.1 & 50(20) \\
 & & & & & \\
 $^{12}$N & $2^-$ & 1.191 & $p_0(s)$ & 0.590 & 124 & 87
         & 118(14) \\
 & $1^-$ & 1.764 & $p_0(s)$ & 1.163 & 1173 & 894 & 750(250) \\
 & $3^-$ & 3.132 & $p_0(d)$ & 2.531 & 250 & 220 & 220(25) \\
 & $1^-$ & 3.449 & $p_0(s)$ & 2.848 & \tablenotemark[1] & & \\
 &       &       & $p_1(s)$ & 0.848 & 403  & 311 & 260(30) \\
 & $2^-$ & 4.048 & $p_0(d)$ & 3.447 & 640 & 290 & \\
 &       &       & $p_0(s)$ & 3.447 & \tablenotemark[1] & & 836(25) \\
 & $4^-$ & 4.237 & $p_0(d)$ & 3.636 & 749 & 610 & 744(25) \\
 & $3^-$ & 5.348 & $p_0(d)$ & 4.747 & 1614 & 55 & \\
 &       &       & $p_1(d)$ & 2.747 & 323  & 225 & 180(23)\\
\tablenotetext[1]{Too unbound for a single-particle $1s_{1/2}$ proton 
resonance.}
\end{tabular}
\label{table4}
\end{table}

\begin{table}
\caption{Parentages for states in the giant-resonance region.}
\begin{tabular}{lccccc}
\hline
  & $1_3^-$ & $1_4^-$ & $1^-_5$ & $2_3^-$ & $2_4^-$ \\
\hline
 ${3\over 2}_1^-\otimes s$ & 0.056 & 0.004 & 0.006 & 0 005 & 0.021 \\
 ${3\over 2}_1^-\otimes d$ & 0.579 & 0.289 & 0.158 & 0.446 & 0.275 \\ 
 ${1\over 2}_1^-\otimes s$ & & 0.057 & & & \\ 
 ${1\over 2}_1^-\otimes d$ & 0.104 & 0.105 & 0.001 & 0.400 & 0.088 \\ 
 ${5\over 2}_1^-\otimes s$ & & & & 0.015 & 0.062 \\ 
 ${5\over 2}_1^-\otimes d$ & 0.127 & 0.030 & 0.027 & 0.042 & 0.010 \\ 
 ${3\over 2}_2^-\otimes s$ & 0.048 & 0.097 & 0.610 & 0.021 & 0.310 \\ 
 ${3\over 2}_2^-\otimes d$ & 0.011 & 0.353 & 0.061 & 0.027 & 0.173 \\ 
 $E_>$              & 0.075 & 0.065 & 0.137 & 0.044 & 0.061 \\ 
 & & & & \\
 $E_x(^{12}$N) & 7.1 & 7.8 & 8.5 & 6.1 & 6.8 \\
\end{tabular}
\end{table}

\begin{table}
\caption[c]{Partial widths for states in the giant-resonance region.
The symbols are defined in Table\ IV and all energies are in MeV. 
The widths are $d$-wave unless otherwise specified. We estimate that
the width for $s$-wave $p_3$ decay from  the $2^-_4$ level is about
0.6 MeV and that the $1^-_5$ level should be rather broad on account
of the same decay channel (see Table V).}
\begin{tabular}{cccccccc}
 & $J^\pi$  & $E_x$ & $N_n$ & $E_p^{expt}$ & $\Gamma^{sp}$
 & $\Gamma^{th}$ & $\Gamma^{tot}$ \\
\hline
 $^{12}$B & $2_3^-$ & 6.52 & $n_0$ & 3.15 & 1.45 & 0.65 & \\
          &       &        & $n_1$ & 1.03 & 0.10 & 0.04 & 0.69 \\
          & $2_4^-$ & 7.27 & $n_0$ & 3.90 & 2.40 & 0.66 & \\
          &       &        & $n_1$ & 1.78 & 0.37 & 0.03 & 0.69 \\
          & $1_3^-$ & 7.56 & $n_0$ & 4.19 & 2.85 & 1.65 & \\
          &       &        & $n_1$ & 2.07 & 0.55 & 0.06 & 1.71 \\
          & $1_4^-$ & 8.34 & $n_0$ & 4.97 & 4.35 & 1.26 & \\
          &       &        & $n_1$ & 2.85 & 1.15 & 0.12 & 1.38 \\
 & & & & & \\
 $^{12}$N & $2_3^-$ & 6.07 & $p_0$ & 5.47 & 2.42 & 1.08 & \\
          &       &        & $p_1$ & 3.47 & 0.65 & 0.26 & 1.34 \\
          & $2_4^-$ & 6.76 & $p_0$ & 6.16 & 3.37 & 0.93 & \\
          &       &        & $p_1$ & 4.16 & 1.13 & 0.10 & 1.03 \\
          & $1_3^-$ & 7.13 & $p_0$ & 6.53 & 4.03 & 2.33 & \\
          &       &        & $p_1$ & 4.53 & 1.42 & 0.15 & 2.48 \\
          & $1_4^-$ & 7.84 & $p_0$ & 7.24 & 5.50 & 1.59 & \\
          &       &        & $p_1$ & 5.24 & 2.16 & 0.23 & 1.82 \\
\end{tabular}
\end{table}

\begin{table}
\caption{Peak cross sections for the giant dipole region from 
$(p,p')$ ($\times 2$) and charge exchange reactions at 100 to 
200 MeV incident energy.}
\begin{tabular}{cccc}
Reaction & Incident energy & Cross section & Reference \\
         &    MeV     &   mb/sr &  \\
\tableline
 $(p,n)$ &  135    & 3.3 & This work \\
 $(p,n)$ &  120    & 4.7 &  \cite{GA84} \\
 $(p,n)$ &  160    & 4.0 & \cite{GA84} \\
 $(p,n)$ &  200    & 3.5 & \cite{GA84} \\
 $(n,p)$ &  ~98    & 5.7 & \cite{OL93} \\
 $(p,p')$ &  156   & 3.7 & \cite{TE93} \\
 $(p,n)$ &  186    & 3.6 & \cite{YA93} \\
 $(n,p)$ &  190    & 3.6 & \cite{YA93} \\
\end{tabular}
\label{table7}
\end{table}

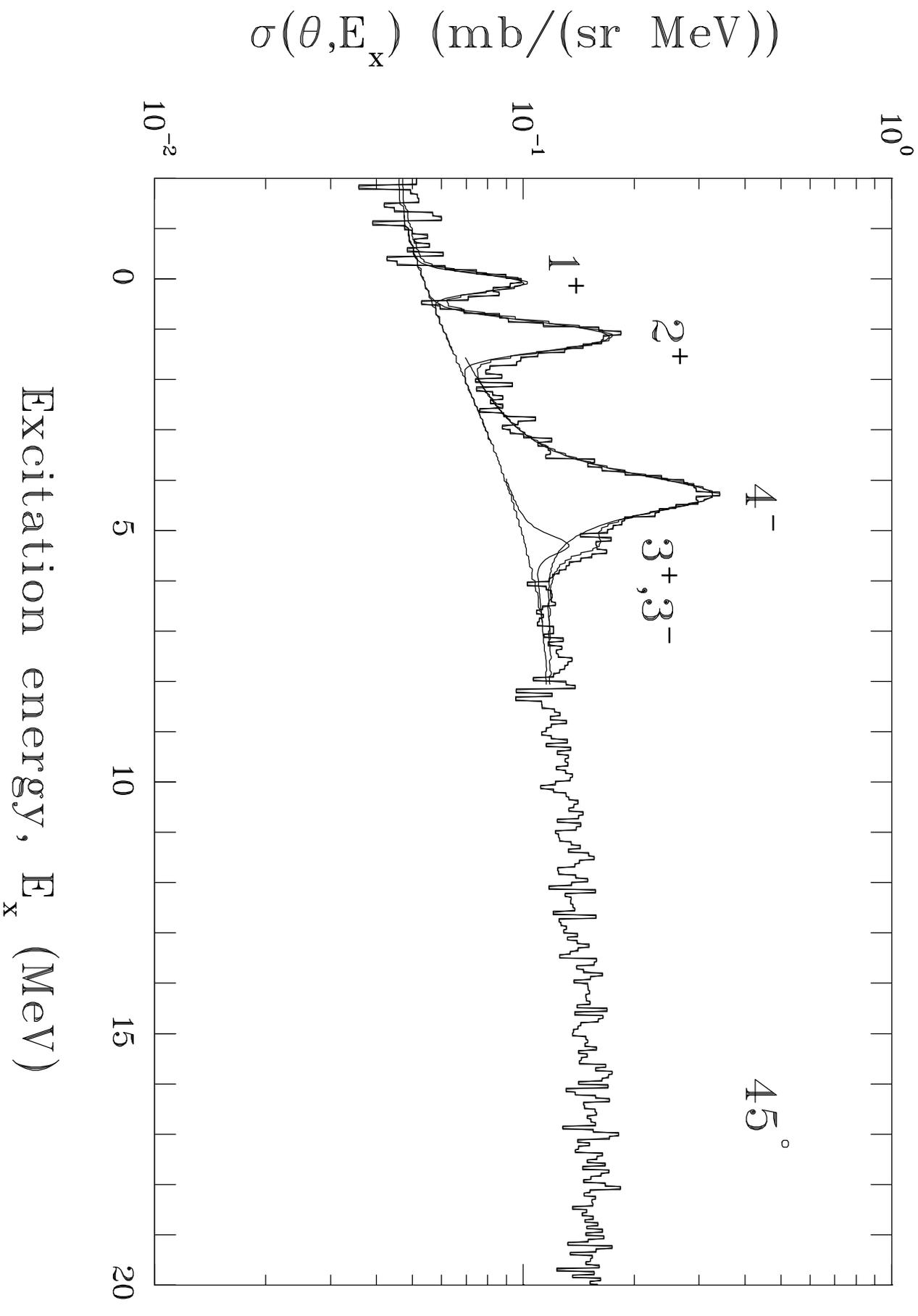
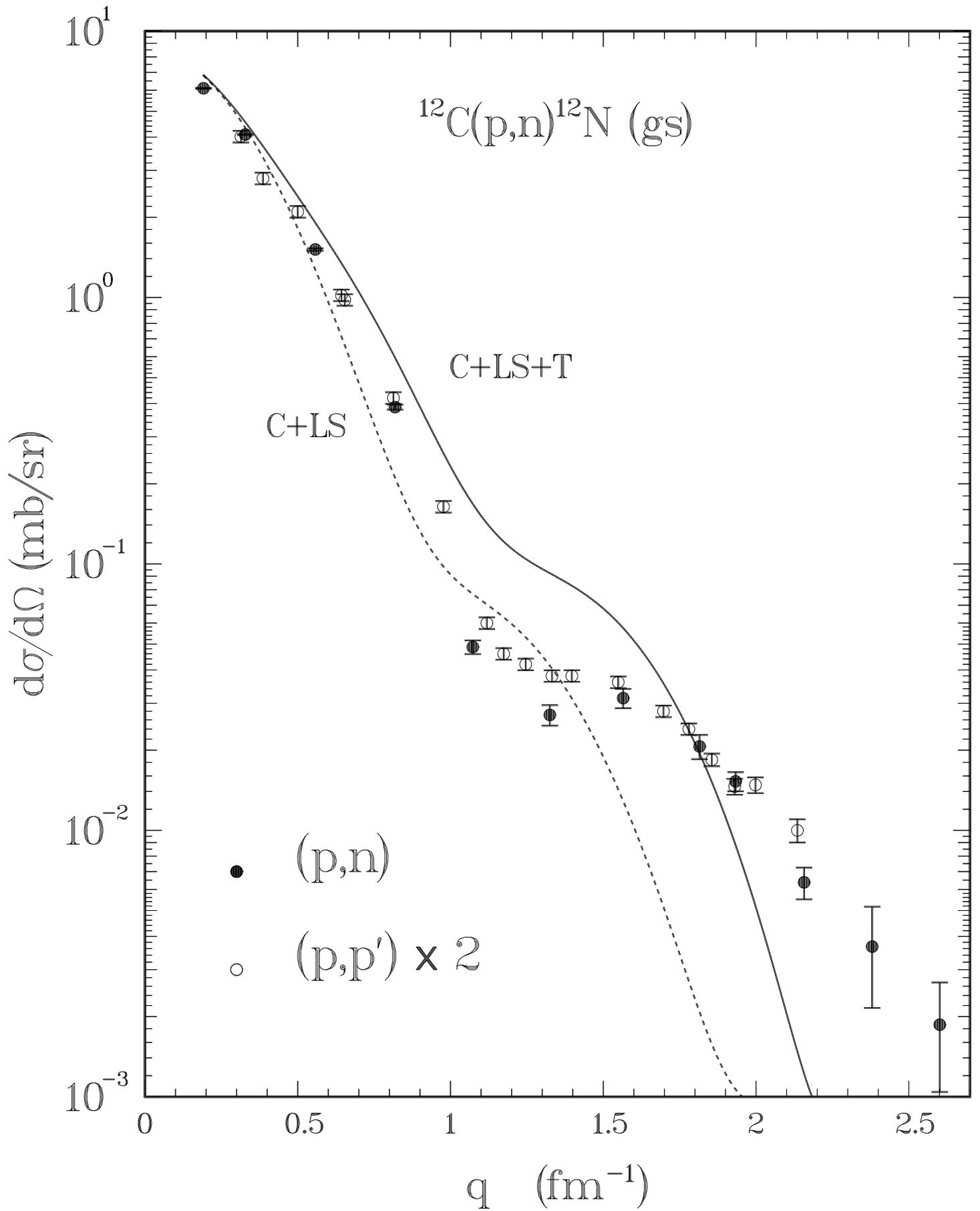
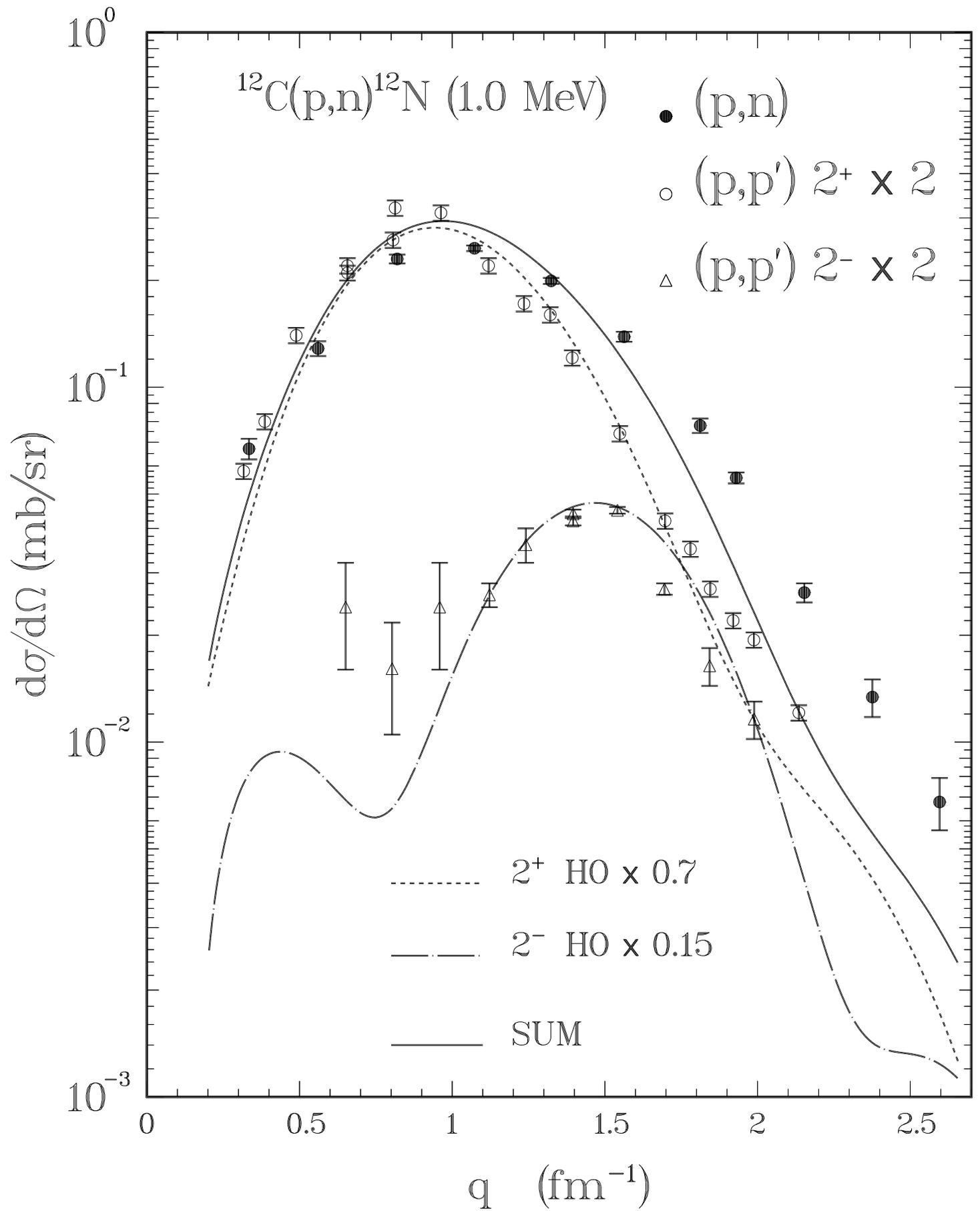
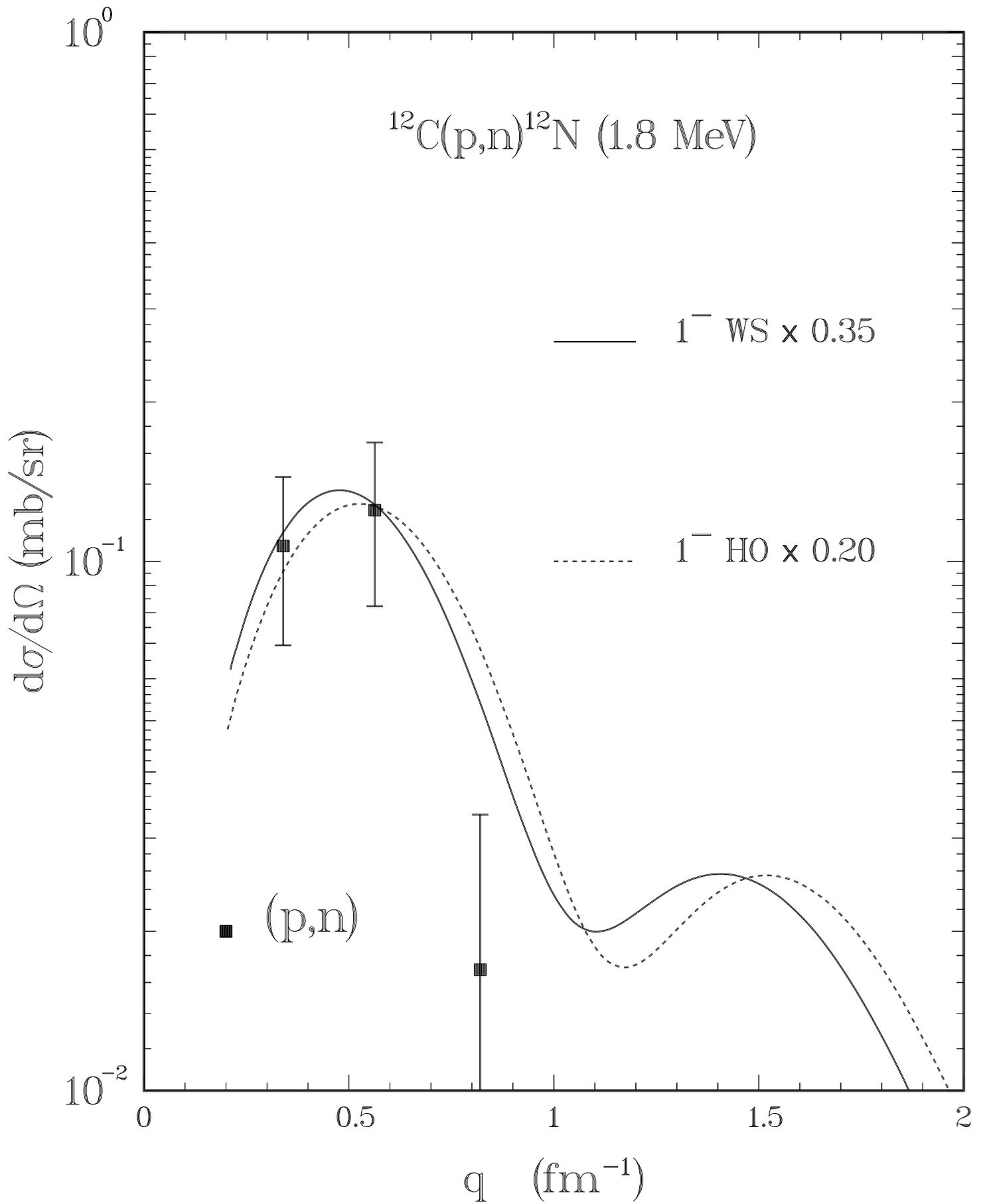
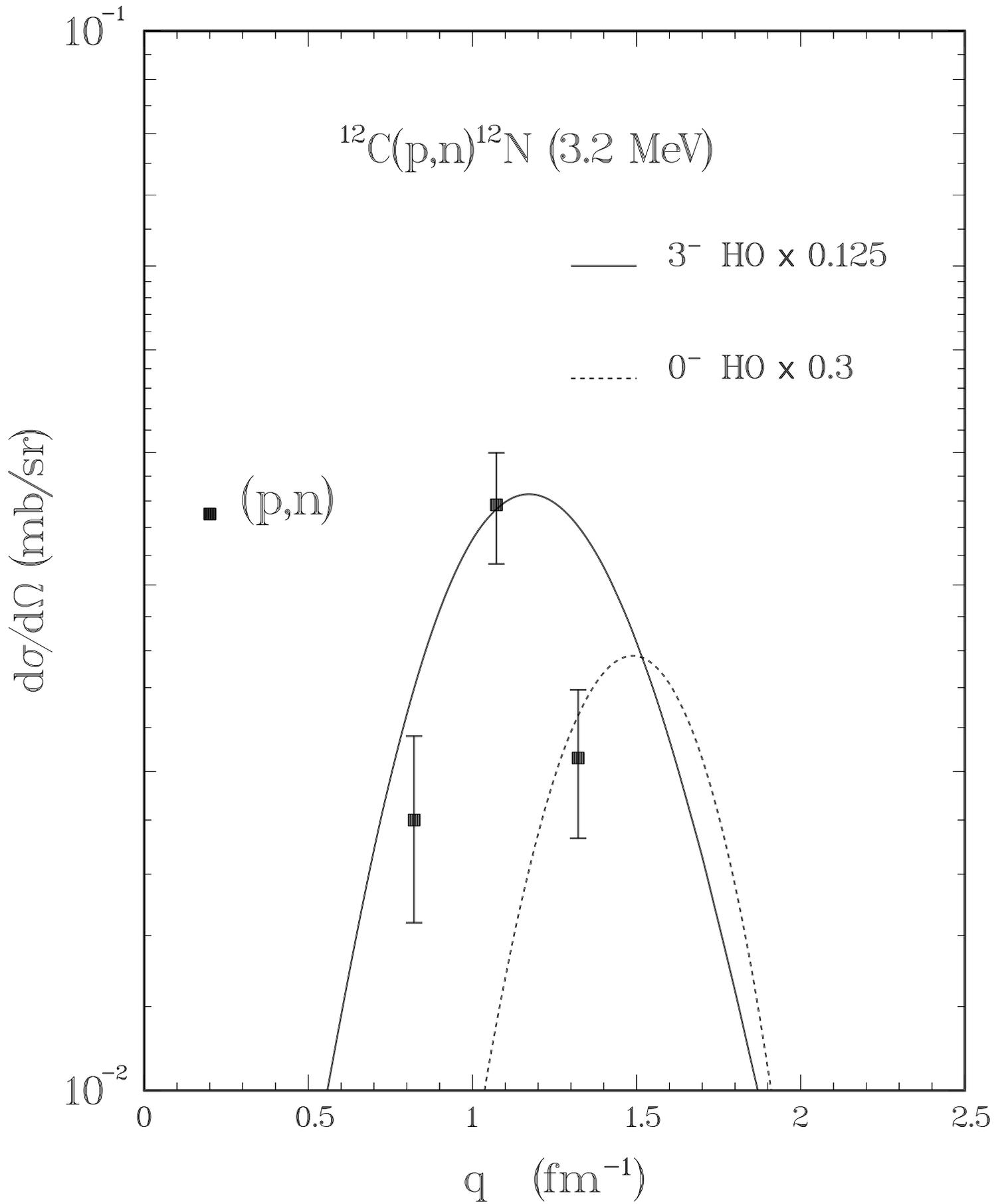
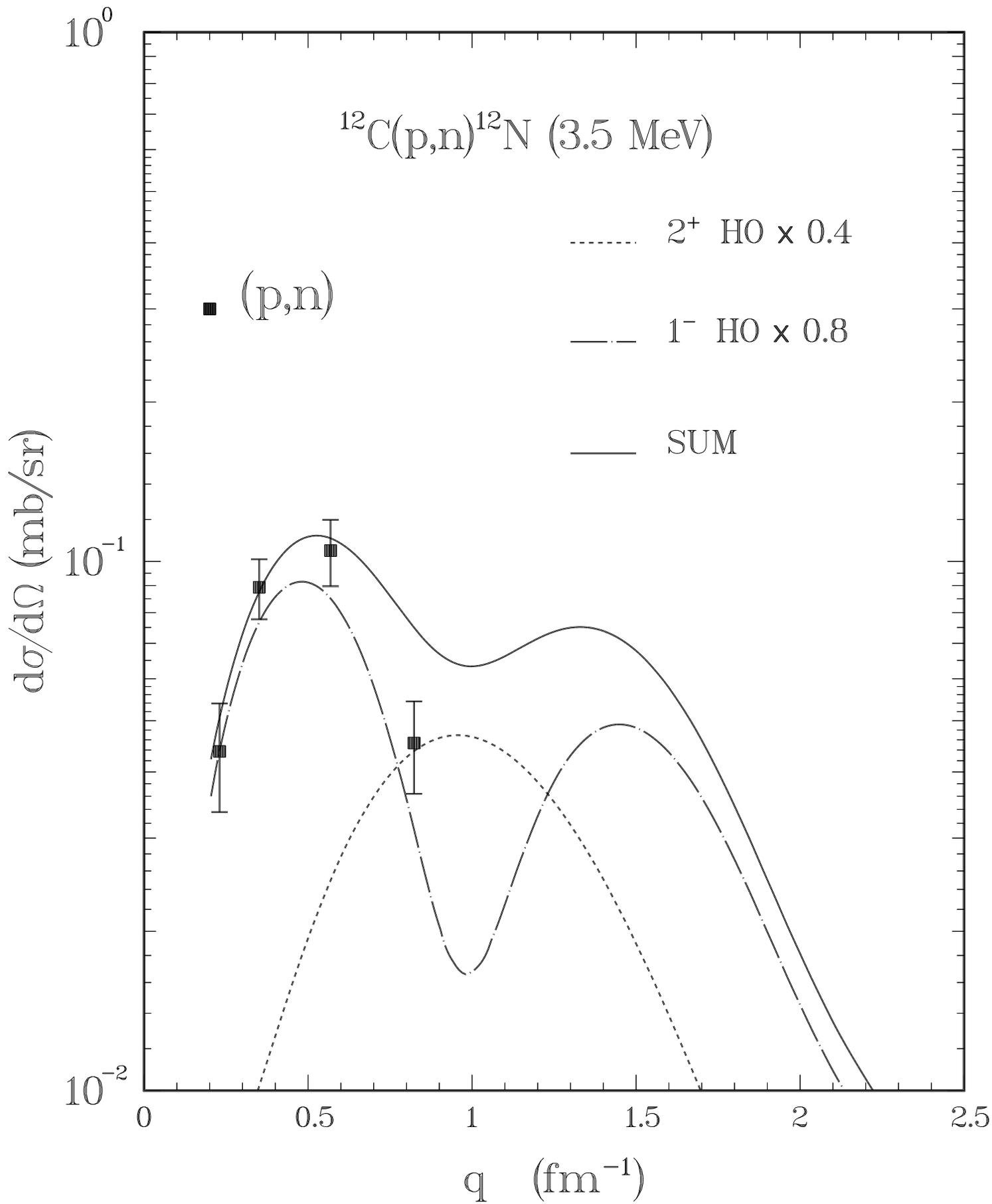
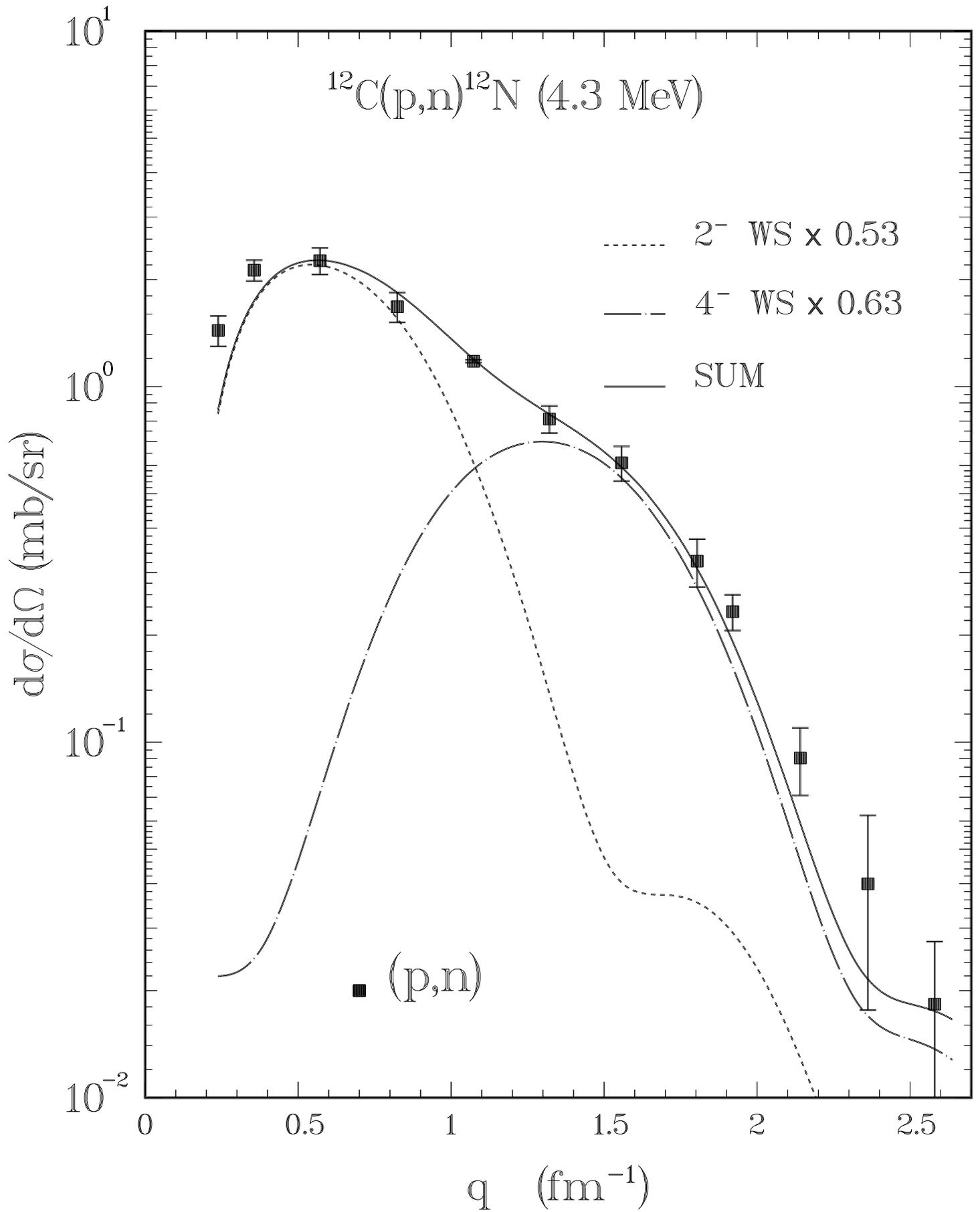
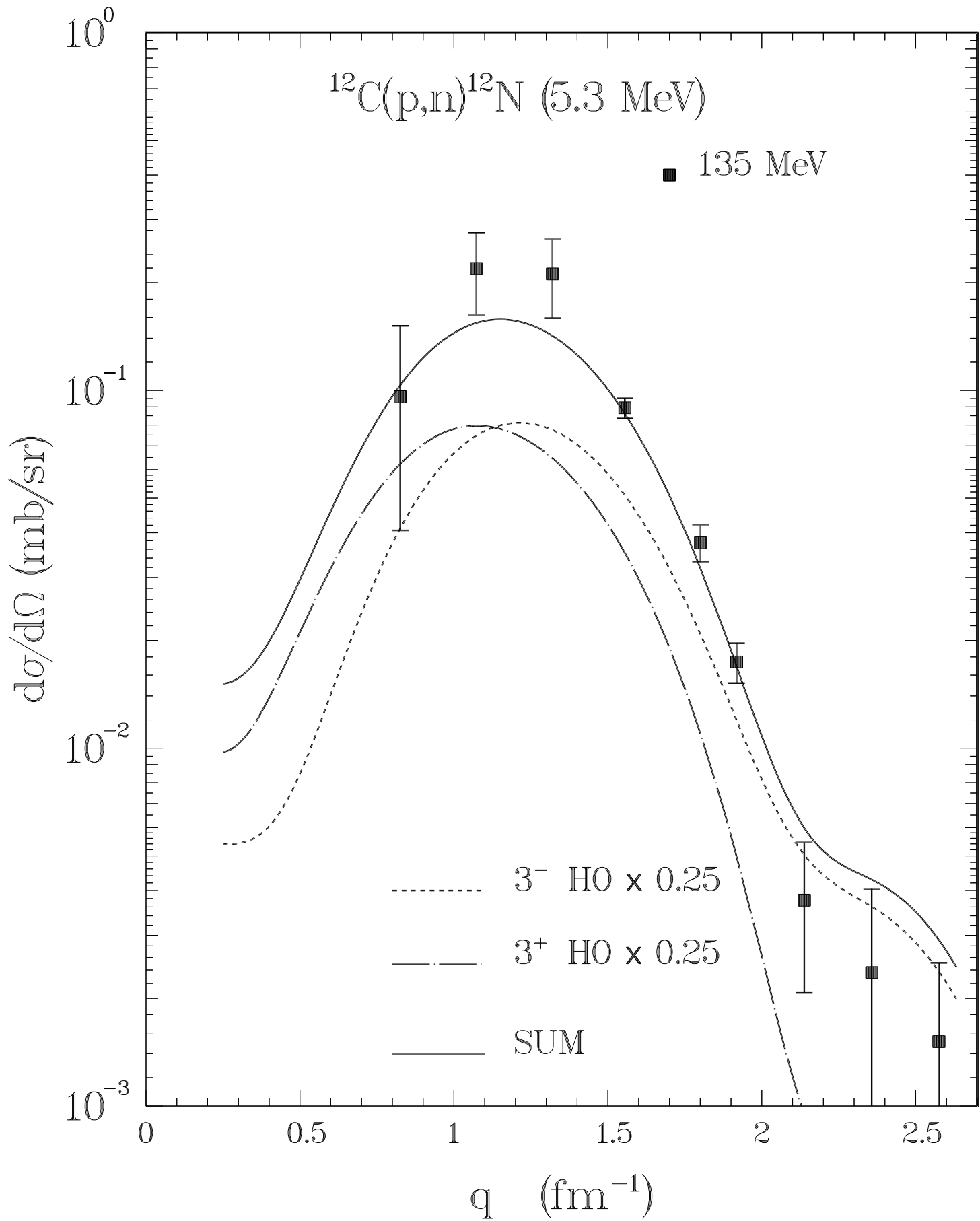
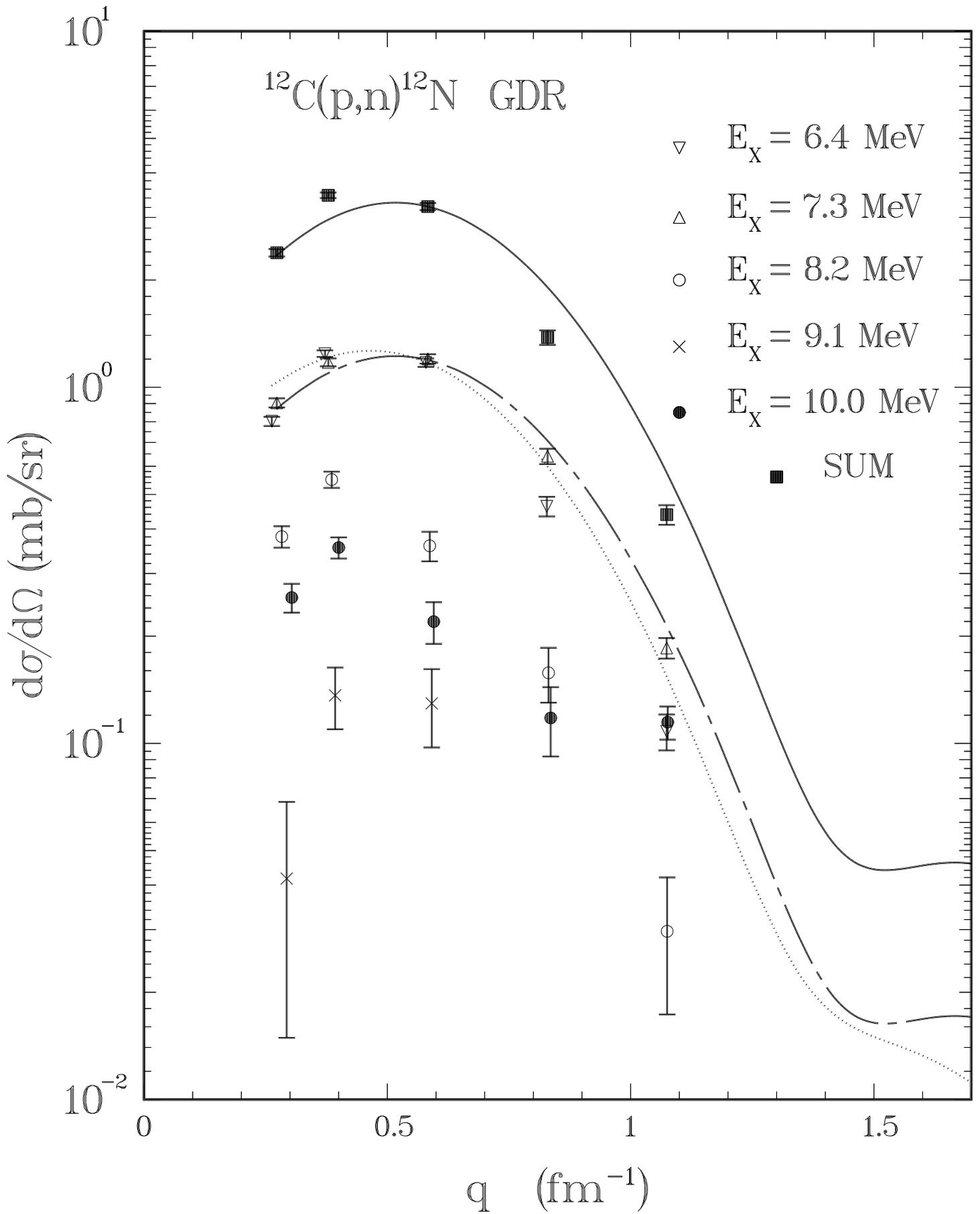
\end{document}